\documentclass[aps,prd,superscriptaddress,showpacs,nofootinbib,eqsecnum,amsfonts,amsmath]{revtex4-1}

\usepackage{epsfig}
\usepackage{longtable}

\def\nn{\nonumber}
\def\beq{\begin{equation}}
\def\eeq{\end{equation}}
\def\be{\begin{equation}}
\def\ee{\end{equation}}
\def\bea{\begin{eqnarray}}
\def\eea{\end{eqnarray}}

\def\ms{\overline{\rm MS}}

\def\t{\tilde }

\def\ga{\gamma}
\def\ds{\displaystyle}
 
\newcommand{\lsim}{\raisebox{-0.13cm}{~\shortstack{$<$ \\[-0.07cm] $\sim$}}~}
\newcommand{\gsim}{\raisebox{-0.13cm}{~\shortstack{$>$ \\[-0.07cm] $\sim$}}~}

\begin{document}

\title{ 
 Renormalization Group Optimized Perturbation Theory at Finite Temperatures
} 

\author{Jean-Lo\"{\i}c Kneur}
\email{jean-loic.kneur@univ-montp2.fr}
\affiliation{Laboratoire Charles Coulomb (L2C), UMR 5221 CNRS-Univ. Montpellier 2, Montpellier, France}

\author{Marcus B. Pinto}
\email{marcus.benghi@ufsc.br} 
\affiliation{Departamento de F\'{\i}sica, Universidade Federal de Santa
  Catarina, 88040-900 Florian\'{o}polis, Santa Catarina, Brazil}
  
\begin{abstract}
 A recently developed variant of the
so-called optimized perturbation theory (OPT), making it perturbatively consistent with 
renormalization group (RG) properties, RGOPT, was shown to drastically improve its convergence for zero
temperature theories. Here the RGOPT adapted to finite temperature is illustrated 
with a detailed evaluation of the two-loop  
pressure for the thermal scalar $ \lambda\phi^4$ field theory. We show that already at the simple one-loop level 
this quantity is exactly scale-invariant by construction and turns out to qualitatively 
reproduce, with a rather simple procedure, results from more sophisticated resummation methods 
at two-loop order, such as the two-particle irreducible approach typically. This lowest order also 
reproduces the exact large-$N$ results of the $O(N)$ model. Although very close in spirit, our RGOPT method and corresponding results differ drastically from 
similar variational approaches, such as the screened perturbation theory or its QCD-version, the  
(resummed) hard thermal loop perturbation theory. 
The latter approaches exhibit a sensibly degrading scale dependence at higher orders, 
which we identify as a consequence of missing RG invariance. In contrast RGOPT gives a considerably reduced 
scale dependence at two-loop level, even for relatively large coupling values $\sqrt{\lambda/24}\sim {\cal O}(1)$, 
making results much more stable as compared with standard perturbation theory,
with expected similar properties for thermal QCD.
\end{abstract}
\pacs{12.38.Lg , 11.10.Wx, 11.10.Gh}

\maketitle

\section{Introduction}
It is a well established fact that evaluations devoted to describe  quantum chromodynamics (QCD) phase transitions need to be done in a nonperturbative 
fashion, such as numerically solving it on the lattice (LQCD)~\cite{aoki}, which nowadays  is considered the 
most reliable way to  tackle the problem at vanishing densities.  Unfortunately, LQCD is still 
plagued by the so-called sign problem~\cite{signpb} which prevents the method to be used to describe the  
transitions  expected to  take place at  lower temperatures and higher densities. 
On the other hand, multi-loop perturbative results for many 
QCD physical  quantities are within  reach so that an appealing alternative would 
be to use them in conjunction with some resummation procedure in order to generate 
nonperturbative results. In this vein, different analytical techniques envisaged to combine 
the easiness of purely perturbative evaluations with nonperturbative optimization/resummation 
procedures have been proposed in the past decades~\cite{Trev}. Some of these methods are  based on a 
reorganization of a given interacting Lagrangian, 
so that it becomes written in terms of an arbitrary mass parameter 
which, for massless theories, also works as an infrared regulator (as in 
hard thermal loop resummations~\cite{HTL,SPTearly}). 
One of these approaches is the so-called screened perturbation theory (SPT)~\cite{SPT,SPT3l}, 
in which the variational  parameter is described by a thermal mass. The SPT was originally 
proposed to describe the thermodynamics of massless scalar theories, 
but it has been later generalized  so that the equation of state of thermal gauge-invariant theories~\cite{HTL}, 
such as QCD, could also be obtained. This gauge invariant generalization known as  hard thermal 
loop (resummed) perturbation theory (HTLpt)~\cite{htlpt1}, has been already used  to calculate QCD thermodynamic 
functions up to three loop order at finite values of the  temperature and chemical 
potential~\cite{HTLPT3loop,HTLPTMU}. 
The SPT and HTLpt are actually conceptually similar to the so-called linear delta expansion (LDE)
and optimized perturbation theory (OPT), developed earlier under various different names~\cite{OPT-LDE,odm,pms} mainly 
in the context of zero-temperature field theories. Within this technique, perturbative 
evaluations are performed using propagators written in terms of 
an arbitrary mass parameter, so that optimized nonperturbative results can be generated by 
requiring the mass parameter to satisfy a variational criterion.
 The two major  problems the above mentioned methods try to solve are the poor convergence 
 and the notoriously bad scale-dependence 
of the standard perturbative series both for the thermal mass and for the pressure 
at higher orders (see e.g. \cite{Trev} for a review): not only the increasing perturbative
orders show no clear sign of stability, 
but the scale-dependence worsens substantially at higher orders,  
at odds with what is intuitively expected for most 
known perturbative series at $T=0$. Part of this bad
behavior is commonly explained~\cite{Trev} by the unavoidable complicated interplay of soft 
and hard thermal contributions. Actually,  the dynamical generation of a 
thermal screening mass $m_D\sim \sqrt{\lambda} T$   influences 
the relevant expansion of physical quantities, such as the pressure, which are then expressed 
in powers of $\sqrt \lambda$ rather than $\lambda$. In this case the predictions are,
a priori, less convergent than for the $T=0$ case, yet most of the interesting thermal physics happens
at rather moderate coupling values, so that one could expect a better behavior. Despite 
the unavoidable $\sqrt\lambda$ ``nonperturbative" dependence, both SPT \cite {SPT,SPT3l} and 
OPT~\cite{OPT3l} applications to hot scalar theory show how these methods indeed improve 
the stability of the predictions when  higher orders in the loop expansion 
are considered. Given the inherent technical  
difficulties associated with the (three loop) evaluation of the QCD pressure for the case of hot and dense 
quark matter, the recent results in \cite{HTLPTMU} represent an impressive achievement. 
However,  the same results exhibit a substantial {\em increasing} 
scale-dependence at increasing two- and three-loop orders, 
even for moderately large coupling values, which remains a surprising issue.
This is  more pronounced for HTLpt in QCD applications at the three loop level \cite{HTLPT3loop,HTLPTMU}. 
While the latter are sometimes remarkably close to lattice results for temperatures down to $T \gsim 2 T_c$,  
for the central renormalization scale choice $\mu\sim 2\pi T$ in the $\ms$-scheme, it appears puzzling that 
a moderate scale variation of a factor 2 dramatically affects 
the pressure and related thermodynamical quantities by relative variations of order 1 or more. 
It is argued~\cite{HTLPTMU} that resumming the logarithmic dependence of HTLpt results may improve this problem, but
as we shall explain below the missing RG invariance properties is more basic
within the SPT/HTLpt approach.\\

Recently, the standard OPT procedure at zero temperature has been modified to incorporate
consistently perturbative renormalization group (RG) 
properties. It was shown to considerably improve the convergence of OPT, as tested 
for the Gross-Neveu (GN) model mass gap~\cite{rgopt1}, 
and further used to determine with a good accuracy 
the basic QCD scale in the $\overline {\rm MS}$ scheme, 
$\Lambda_{\overline {\rm MS}}$, and corresponding value of the 
strong coupling ${\bar \alpha}_S$\cite{rgopt_Lam,rgopt_alphas}. Very recently the same method has been used to
estimate~\cite{rgopt_qq} the QCD chiral quark condensate. 
Here, we extend the construction to the case of a scalar theory with quartic interaction to show how this 
RG improved OPT (RGOPT) can easily cope with the introduction of 
control parameters such as the temperature,  developing in  detail the construction presented very recently 
in ~\cite{rgopt_phi4_lett}. Using this textbook example 
we aim to illustrate how the RGOPT takes care of the scale dependence problems of thermal
theories, those being more clearly visible within the SPT \cite {SPT,SPT3l} and HTLpt higher 
loop results~\cite{htlpt1,HTLPT3loop,HTLPTMU}, but also present in other resummation approaches. \\
In a nutshell, leaving aside technical details, our basic observation is that the arbitrary variational mass 
introduced in the SPT/OPT context can (and should) be treated as any ``proper" mass, from the RG viewpoint. 
In particular, this means that  
it involves its own standard anomalous dimension, 
which is unrelated to the coupling $\beta$ function, and should be incorporated consistently within RG properties. 
 This brings crucial
consequences already on the form of an explicitly RG-invariant pressure, prior to any 
further improved/resummed perturbation approaches. 
It also more clearly separates the hard and soft modes from a RG/scale
dependence perspective, at least in an intermediate stage of the calculations. In the more standard approach  
both contributions are mixed up since all relevant quantities are expressed as a function of the coupling 
and temperature only.
That being accepted, one realizes that most of 
the observed worsening scale-dependence at higher orders is actually due to a manifest failure of RG
invariance for a massive theory. This is at least particularly transparent  
 in the $\ms$ scheme largely used within  SPT/HTLpt methods, 
 where we show that minimally
 subtracting the vacuum energy divergences, without finite vacuum energy contributions, 
 explicitly misses RG invariance. \\
 In addition, even when starting from an explicitly RG invariant
 perturbative expression, RG invariance is generally lost
 as a consequence of the standard (linear) modification of perturbative expansion implied in OPT/SPT,
 a fact that has been seldom appreciated in the relevant literature so far. However, RG properties 
can easily be restored by a consistent use of renormalization
group properties for a massive theory, automatically incorporated in the RGOPT approach~\cite{rgopt_alphas}, 
whereby a drastically improved scale-dependence follows naturally. This opens up the possibility of rather simply
exploiting many non-trivial SPT/HTLpt results performed up to three-loop order so far, by improving substantially
their scale independence after applying appropriate RGOPT adaptations.\\ 
Incidentally, other resummation approaches, like typically 
the so-called ``two-loop-$\Phi$-derivable" approach, 
(related to the two-particle-irreducible (2PI) method)~\cite{2PItadpole,2PI} are   
 manifestly scale invariant when applied to the scalar model 
 in a certain approximation. 
 In fact one observes 
 interesting analogies (but also important differences) between the 
 first non-trival (two-loop) 2PI order and 
 our {\em one-loop} RGOPT results, as will be elaborated in more detail below. 
 However, as it happens the scale-invariance of the above-mentioned 2PI results~\cite{2PItadpole,2PI}
 is essentially due to a renormalization that appears peculiar to the scalar model, 
 inspired by the properties of the $O(N)$ symmetric theory 
 in the large-$N$ limit~\cite{phi4N} (for which indeed the two-loop-$\Phi$-derivable 2PI results become exact). 
 Thus, to the best of our understanding, this scale invariance appears somewhat accidental and difficult 
 to translate to higher orders and to QCD. Incidentally, the 2PI approach has been pushed even to three-loop order 
 for the $\phi^4$ theory in Ref.~\cite{2PI3loop}, with remarkably stable results with respect to two-loop order,  
  but the method becomes more involved, and a definite scale-dependence reappears, although much more moderate than 
  in the three-loop order SPT and HTLpt cases. 
 Other approaches like the nonperturbative renormalization group (NPRG)~\cite{NPRG},
 should be RG invariant by construction. But solving the  relevant NPRG   
 equations for thermal QCD beyond approximative truncation schemes 
 becomes very involved in practice.
  We also remark  that our approach shares some 
qualitative features with the ideas invoked and the framework developed 
very recently in Ref. \cite{Blaizot-Wschebor}, in which the authors explored some 
form of massive renormalization scheme and its consequences. However, the  
RGOPT differs substantially in its approach as it
 incorporates by construction \cite{rgopt_alphas,rgopt_phi4_lett} 
all relevant RG properties systematically order by order, 
relying basically on standard perturbation theory. Being largely based on (but not limited to) 
$\ms$-scheme results, the RGOPT can easily be extended to any higher order
calculations performed in the framework of different models, when already available, including thermal QCD.
Therefore,  it appears to us that the method presented here is conceptually  simpler than  other more 
sophisticated resummation approaches mentioned above. \\

The paper is organized as follows. In the next section we 
quickly review some basic results of thermal scalar theory for the 
free energy at the relevant two-loop level. Then, in section 3 we address in some details 
the RG (non) invariance issue in the massive case, paying special attention  
to the $\ms$ scheme largely used in the SPT (or the similar HTLpt) method. In the same section we also 
explain how to restore the perturbative RG invariance at arbitrary orders in a simple fashion. 
Next, the resummation by optimized perturbation (OPT),
with the crucial modification to maintain its perturbative RG invariance, RGOPT, 
is discussed in rather general terms in sections 4 and 5 respectively. The method  
is then illustrated in details by evaluating the free energy of a hot scalar field theory 
at one- and two-loop level in sections 6 and 7. We emphasize that all 
the construction developed in Secs. 3 to 7 for the $\phi^4$ model (for which we also  
briefly consider the large-$N$ case) is actually more general, most of it being straightforwardly 
applicable to thermal QCD. We occasionally mention some 
properties anticipated to be similar (or eventually different) in the QCD case.
 Finally, our conclusions are  presented
in section 8, and one appendix deals with some technical details on the RG-invariant
construction of counterterms.
\section{Finite temperature scalar field theory}
We consider, as a starting point, the massive scalar field theory described by  the Lagrangian density
\be
{\cal L} = \frac{1}{2} \partial_\mu \phi \partial^\mu \phi -\frac{m^2}{2}\phi^2 -\frac{\lambda}{4!} \phi^4\,\,;
\label{Lphi4}
\ee
where we introduce a (yet unspecified) generic mass term ($m$) which can be thought as a thermal mass
generated by higher perturbative orders in an originally  massless theory.
One may then evaluate the free  energy using known results from ordinary perturbation theory 
for the massive case.
Then, up to the two-loop 
level the basic expression of the (bare) free energy is formally~\cite{kapusta,SPT3l,Trev}:
\be
{\cal F}_0 = \frac{1}{2}
\int_p \ln (p^2+m^2)  +
\frac{\lambda}{8} \left(\int_p \frac{1}{p^2+m^2} \right)^2 +{\cal F}_0^{\rm ct}\;,
\label{2lbasic}
\ee
where the temperature is introduced via Matsubara's imaginary time formalism 
($p^2\equiv \omega^2_n+\bf p^2$ with the bosonic Matsubara
$\omega_n=2\pi n T$) and  we have also defined  
\be
\int_p \equiv \left (e^{\gamma_E} \frac{\mu^2}{4\pi} \right )^{\epsilon}\,T \sum_n \int \frac{d^{D-1} p}{(2\pi)^{D-1}} \,.  
\ee
The divergent integrals are regulated using dimensional regularization techniques 
with $D=4-2\epsilon$ while renormalization is carried out in the  $\ms$-scheme. The term ${\cal F}_0^{\rm ct}$ 
represents all the relevant  counterterms contributions   to $O(\lambda)$ (see Ref. \cite {SPT3l} for details). 
After all the mass, coupling, and vacuum energy
counterterms have been consistently introduced 
to cancel the original divergences,
one obtains  the ($\ms$-scheme) renormalized free energy~\cite{SPT3l,Trev}:
\be
 (4\pi)^2 {\cal F}_0 = {\cal E}_0 -\frac{1}{8} m^4 \left [3+2\ln \left ( \frac{\mu^2}{m^2} \right ) \right ] 
-\frac{1}{2} T^4 J_0\left (\frac{m}{T}\right ) + \frac{1}{8}\frac{\lambda}{16\pi^2} \left \{ \left [\ln \left  (\frac{ \mu^2}{m^2}\right )+1\right ] m^2 
-T^2 J_1\left (\frac{m}{T}\right ) \right \}^2 \,,
\label{F02l}
\ee
where we have explicitly separated the thermal and non-thermal contributions for later convenience.
Here and in all related renormalized expressions below, $\mu$ represents the arbitrary 
renormalization scale introduced by dimensional regularization 
in the $\ms$-scheme, and $\lambda\equiv \lambda(\mu)$. Note carefully in Eq.(\ref{F02l}) that  ${\cal E}_0$ represents a possible 
{\em finite} vacuum energy term which is usually ignored, i.e. minimally set to zero in the (thermal)
literature~\cite{SPT3l}. However, within our approach this 
quantity is necessarily non-zero and plays a crucial role as will become clear in the sequel.\\ 
The standard (dimensionless) thermal integrals appearing in Eq.(\ref{F02l}) are given by
\be
J_n(x) = 4 \frac{\Gamma[1/2]}{\Gamma[5/2-n]} \: \int_0^\infty dt \frac{t^{4-2n}}{\sqrt{t^2+x^2}}\:
\frac{1}{e^{\sqrt{t^2+x^2}}-1} \,,
\ee
 where  $t=p/T$ and $x=m/T$. Different integrals can be easily related by employing derivatives such as
\be
J_{n+1}(x) = -\frac{1}{2x} \frac{\partial}{\partial x} J_n(x) \,.
\ee
Also, a high-$T$ expansion such as 
\be
J_0(x) \simeq \frac{16}{45} \pi^4 -4\frac{\pi^2}{3} x^2+ 8\frac{\pi}{3} x^3 +x^4 
\left [\ln \left (\frac{x}{4\pi}\right )+\gamma_E -\frac{3}{4} \right ]  +{\cal O}(x^6)\,,
\label{J0exp}
\ee
is often useful since it represents a rather good approximation as long as $x \lsim 1$, i.e., $T$ larger than $m$.\\  
Finally, for later comparison we recall that the thermal screening mass ($m_D$), defined~\cite{mdeb} 
by the pole of the (static) propagator,
is obtained for weak coupling for the massless theory as a perturbative series which to lowest orders 
reads~\cite{mdeb}:
\be
\frac{m^2_D}{T^2} = \frac{\lambda}{24}\left \{1-\sqrt 6 \left (\frac{\lambda}{16\pi^2}\right )^{1/2}
-\left (\frac{\lambda}{16\pi^2}\right )\left [3\ln \frac{\mu}{2\pi T}-2\ln\frac{\lambda}{16\pi^2}-6.4341 \right ] +{\cal O}(\lambda^{3/2}) \right \}\,.
\label{mD1}
\ee
\section{RG invariant Free Energy in Massive Renormalization Schemes}
We now discuss the lack of RG invariance when ${\cal E}_0$ is minimally set to zero 
in Eq.(\ref {F02l}). Remark first that to obtain (\ref{F02l}), calculations have been performed with an 
arbitrary mass $m$ in the dressed propagators (mainly to subsequently treat it variationally 
in the OPT/SPT approach), with no prejudice at this stage that it should be a thermal mass 
of order $m^2 \sim \lambda T^2$ in the actually massless theory. 
Thus from the RG viewpoint, everything in (\ref{F02l}) should behave
as if it was a genuine massive theory, in particular the mass should have its standard anomalous dimension. 
Recall that the (homogeneous) RG operator acting on a physical quantity with mass dependence, 
such as the free energy in the present case, is defined as
\be
\mu\frac{d}{d\,\mu} =
\mu\frac{\partial}{\partial\mu}+\beta(\lambda)\frac{\partial}{\partial \lambda}
+\gamma_m(\lambda)\,m
 \frac{\partial}{\partial m}\;,
 \label{RGop}
\ee 
where our normalization for the $\beta$ function is
\be
\beta(\lambda)\equiv \frac{d\lambda}{d\ln\mu} = b_0 \lambda^2 
+b_1 \lambda^3 +\cdots
\label{beta}
\ee
 while  for the anomalous mass dimension it is given by 
\be
\gamma_m(\lambda) \equiv \frac{d \ln m}{d\ln\mu}= \gamma_0 \lambda +\gamma_1 \lambda^2 +\cdots
\label{gamma}
\ee
with~\cite{RGphi4loop} 
\be
(4\pi)^2 b_0 = 3; \;\; (4\pi)^2 \gamma_0 = \frac{1}{2};\;\;\; (4\pi)^4 b_1 = -\frac{17}{3}\;,
 (4\pi)^4 \gamma_1 = -\frac{5}{12}\;.
 \label{rgcoeff}
\ee
 It is easy to see that the renormalized expression, Eq.~(\ref{F02l}), 
 requires a finite ${\cal E}_0$ contribution to be RG-invariant, as one can readily see by 
 considering the one-loop term which has an explicit $\ln \mu$ dependence. Thus, acting with the RG operator, 
 Eq.~(\ref{RGop}), on the RHS of Eq.~(\ref{F02l}) 
gives a non-zero contribution of order ${\cal O}(1)$: $-(1/2) m^4$, which is not compensated
by terms in Eq.~(\ref{RGop}) coming from the lowest orders in $\beta(\lambda)\, \partial_\lambda$ or 
$m \gamma_m(\lambda)\partial_m \propto \lambda\, m^4$,  
those being  at least of next order ${\cal O}(\lambda)$. This is a manifestation of the fact that (perturbative) 
RG invariance generally occurs from cancellations between terms coming from RG coefficients 
at order $\lambda^k$ and the explicit $\mu$ dependence at the next order $\lambda^{k+1}$. 
This can also be understood alternatively  
by considering solely the original bare contribution to the free energy: although the latter 
only depends on the RG-invariant bare mass and coupling $m_0, \lambda_0$ (and on $2\epsilon =4-D$ 
in dimensional regularization), its {\em finite} part  
is not a priori {\em separately} RG invariant. In other words 
for a massive theory the $T$-independent vacuum energy divergences cannot be absorbed by an arbitrary 
redefinition of the vacuum energy without spoiling RG-invariance. Now, as we recall below the 
subtraction needed to recover RG invariance is perturbatively well-defined and easy to construct order by order. 
The vacuum energy gets its own anomalous dimension which, within 
dimensional regularization, is essentially 
determined  by the coefficients of the poles 
in $2\epsilon=4-D$, stemming from the remaining divergences once the mass and the coupling have been 
properly renormalized. 
This procedure had been exploited in an earlier application of the 
OPT to evaluate the vacuum energy of the Gross-Neveu (massive) model \cite{gn2} and 
then extended to the QCD case~\cite{qcd1,qcd2}.
Similarly, a well-known related result is that the
Coleman-Weinberg effective potential for a general massive theory is not RG invariant 
without {\em finite} ``vacuum energy'' terms independent of the fields, as was originally 
carried out in Ref. \cite{RGvacen} and in the $\ms$-scheme  
in the context of RG-improvements of the effective potential \cite{RGinvMS}.
Indeed, for the $O(N)$ $\phi^4$ model, the vacuum energy anomalous dimension has
even been computed up to four- and five-loop order in Ref. \cite{vacanom_kastening}.

However, to the best of our knowledge, 
this point appears to have been overlooked in the context of thermal theories. 
In applications of improved/resummed massive perturbation schemes 
such as SPT \cite{SPT},  HTLpt \cite{HTLPT3loop}, and the standard OPT  \cite{OPT3l}
the calculations are mostly performed within the $\ms$-scheme and the $T=0$ 
vacuum energy divergence is minimally cancelled out by appropriate (zero point) counterterms but 
missing out those extra finite subtractions required by RG properties. 
 In fact, as far as the purely perturbative massless theory is concerned,
the only mass is actually a thermal mass: $m^2_{th}\sim \lambda T^2$, so that the lack of RG invariance
pointed out above is rather postponed to higher (three-loop) perturbative order $\lambda^2$, where it plainly resurfaces.
Within the SPT, or the similar HTLpt, approaches the variational mass parameter is similarly perturbatively
of order $m^2\sim \lambda T^2$. It is thus not surprising that the scale dependence observed within  SPT/HTLpt results 
appears to worsen at higher orders \cite{SPT3l,HTLPT3loop}. But more generally one wishes to use the nonperturbative mass gap resulting
from such variational approaches possibly beyond standard perturbation for moderately large coupling values, as can be relevant 
near a critical temperature. Thus, the lack of RG invariance appears more serious
since as we recall in next section (see Eq.~(\ref{subst1}),  
the variational procedure formally treats the mass to be of the same perturbative order 
as the lowest order considered contributions, like e.g. the ``hard" thermal 
one-loop contributions of order $\sim \lambda^0\, T^4$.
 Moreover, in the standard procedure one makes the arbitrary renormalization scale $\mu$ 
effectively temperature dependent by choosing $\mu \sim 2\pi T$, such as to avoid large $\ln \mu/(2\pi T)$ contributions coming from the remnant scale-dependence.
In this way the pressure can be studied as a function of $T/T_c$ in QCD applications, where $T_c$ is related to the basic QCD $\Lambda_{QCD}$, {\it e.g.}
in the $\ms$-scheme.
But if the scale dependence appears not much reliable at higher loop orders, 
one may question as well the reliability of the corresponding $T/T_c$ dependence of the pressure, 
even for the well-motivated central $\mu=2\pi T$ prescription.\\
While those issues in $\ms$ or related schemes 
may perhaps not explain at once all the problems that thermal theories face with 
perturbative expansions at increasing coupling, a part of those problems 
are likely to be reduced if one adopts from the beginning a prescription fully consistent with 
RG properties. This problem appears partly circumvented (but are
actually rather delayed to higher orders) in thermal perturbative calculations 
performed in some other renormalization schemes, where the zero-point energy, ${\cal F}_0(T=0)$, 
is subtracted for convenience prior to any subsequent calculations. Indeed, subtracting 
the $T=0$ contribution from Eq.~(\ref{F02l}) 
washes out all the first $\mu$-dependent terms, making scale-independence (trivially) satisfied 
at one-loop order, and (less trivially) at the two loop level as well. But then the one-loop
result becomes also trivial, with the only left contribution being the pure thermal, third term in the 
RHS of Eq.(\ref{F02l}). So, there are 
 no possible optimized solutions of the OPT/SPT/HTLpt form, which can only  
 be obtained at the two-loop level. 
 Moreover, applying subsequently the standard SPT/OPT procedure anyway 
spoils RG invariance. In any case it is most convenient to have a prescription
generically valid both for zero and finite temperatures, so that subtracting $T=0$ contributions 
is not satisfactory for a more general framework. 
Accordingly, the subtraction procedure we will consider next only depends on $T=0$ contributions
but is generically valid also for $T\ne 0$.
Moreover, a remarkable consequence is that the subsequent mass optimization, as implied by 
RGOPT, will give 
a nontrivial solution already at the one-loop order, and 
very similar to what is normally obtained at two-loop order with
the other mentioned resummation schemes (SPT/OPT, HTLPT, 2PI,...), as we will examine in detail.\\

Following Refs. \cite{gn2,qcd1,qcd2,rgopt_Lam,rgopt_alphas} the easiest way to construct an RG-invariant
finite vacuum energy is to determine ${\cal E}_0$ order by order as a perturbative series 
from the reminder of acting with Eq.~(\ref{RGop})
on the non RG-invariant finite part of Eq.~(\ref{F02l}):
\be
\mu \frac{d}{d\mu}{\rm {\cal E}_0}(\lambda,m) \equiv -{\rm Remnant}(\lambda,m)= -\mu \frac{d}{d\mu}
[{\cal F}_0({\rm {\cal E}_0\equiv 0})|_{\rm finite}]\;
\label{rganom}
\ee
where the RHS of (\ref{rganom}) thus defines the anomalous dimension of the vacuum energy.
As above mentioned, it is easy to see that, as a perturbative series, $ {\cal E}_0$ 
has the convenient form in $\ms$ or similar schemes 
\be
{\rm {\cal E}_0}(\lambda,m) = -\frac{m^4}{\lambda} \sum_{k\ge 0} s_k  \lambda^{k}\;,
\label{subgen}
\ee
where the constant coefficients $s_k$ are perturbatively determined
order by order, being essentially determined by the coefficients
of the (single) powers of $\ln \mu$ term at order $k+1$ 
(or equivalently by the single poles in $1/\epsilon$ of the unrenormalized 
expression)~\cite{qcd1,rgopt_alphas}.
This procedure leaves non RG-invariant remnant terms 
of perturbative higher orders to be cared for similarly once higher order terms are considered.  
The apparently odd divergent behavior for $\lambda\to 0$ of this first order term is actually not a problem since,
as we will see explicitly, it completely disappears from the final results. \\ 
We stress that all the previous considerations,  being only dependent on the renormalization procedure, 
do not depend on the thermal contribution 
so that, at arbitrary perturbative orders, the subtraction function represented by Eq.(\ref{subgen}) 
can be determined simply from the $T=0$ contributions only~\footnote{However, when the $T=0$ and the $T\ne 0$
contributions are not explicitly separated, like in the case of two- and three-loop HTLpt \cite{HTLPT3loop}, 
due to the systematic $m/T$ expansion making such involved calculations tractable, caution will be needed to expand 
at a sufficient order in $m/T$ so as to get all the relevant terms of the same perturbative order
to construct the corresponding subtractions in Eq.~(\ref{subgen}).}. Note that Eq.~(\ref{subgen}) is not the only possible
subtraction form in general, but a very convenient one in $\ms$ or related schemes to proceed systematically at higher orders.
Moreover, it is particularly convenient once introducing below the RGOPT modification of perturbation, since
the $1/\lambda$ term will be responsible for a non trivial RGOPT solution already at one-loop order. \\

Explicitly, one  finds the 
RG-invariant form  of Eq.~(\ref{F02l}) up to two-loop order with a non-trivial ${\cal E}_0$ given by
\be
{\cal E}_0 = -m^4 \left [\frac{s_0}{\lambda}+s_1 +s_2 \lambda +{\cal O}(\lambda^2)\right ]\,.
\label{sub}
\ee
 After some algebra one obtains
\bea
& s_0 =\frac{1}{2(b_0-4\gamma_0)} = 8\pi^2\;;\;\;\;
s_1=\frac{(b_1-4\gamma_1)}{8\gamma_0\,(b_0-4\gamma_0)} = -1\; ; \nonumber \\
& s_2 = \frac{96\pi^2\,(b_0-128\pi^2\left((1+4s_1)\gamma_1-s_0(b_2-4\gamma_2)\right)-41}
{12288\,\pi^4(b_0+4\gamma_0)} = \frac{23+36\zeta[3]}{480\,\pi^2}\simeq 0.01399 \;,
\label{s0s1}
\eea
where the explicit RG dependence in the intermediate terms emphasizes 
the more general form of these results, while the last terms are specific to the $N=1$ $\phi^4$ theory. To derive
$s_2$ according to the previous discussion we had to use the ($T=0$) $\ln \mu$ coefficient at three-loop 
order given e.g. in Ref. \cite{SPT3l}. \\
One may equivalently  derive  the finite subtraction in Eq.~(\ref{sub}) 
in an alternative manner by RG invariance considerations solely 
on the bare expression of the free energy. For completeness, this is  presented in the appendix. 
Instead of minimally subtracting the bare vacuum energy divergence,  
an RG-invariant counterterm can be added to cancel the remnant divergences, and necessarily
incorporates also the same finite subtraction terms in 
Eq.~(\ref{s0s1}). As a non-trivial crosscheck of our calculation, let us note that now acting with 
the RG operator, Eq.~(\ref{rganom}), on  
the results given by Eqs. (\ref{sub}) and (\ref{s0s1}) one recovers, for $N=1$, the results up to $\lambda^3$ 
of the anomalous dimension $\beta_v(\lambda)$ (with $\mu d {\rm {\cal E}_0}/d\mu \equiv 2\,m^4 \beta_v(\lambda)$) which has been 
calculated up to four and five loops 
for arbitrary $N$ in  Ref.~\cite{vacanom_kastening}.
(Actually we could have used directly the results in \cite{vacanom_kastening} in the present scalar model 
case to derive the $s_k$ in Eq.~(\ref{s0s1}), 
but the above derivation using the basic available perturbative expressions shows precisely how to proceed for an arbitrary
theory, where the vacuum energy anomalous dimension may not always be explicitly available.) \\
One sees that the subtraction with $s_k$, explicitly depending
on RG coefficients, incorporates a non-trivial RG-dependence already at first (one-loop) order, only
depending on already known one-loop standard RG coefficients. This result has important
consequences for the subsequent OPT application. Now, there is a subtlety at this stage: while the 
$s_k$ subtraction terms are strictly necessary to recover RG invariance at order $\lambda^k$, i.e. 
up to neglected $\lambda^{k+1}$ terms, they enter
the free energy expression at order $\lambda^{k-1}$ as Eq.~(\ref{subgen}) indicates. For instance only $s_0$ is needed
to recover RG-invariance at one-loop ${\cal O}(1)$, but the next term $s_1$ is of order ${\cal O}(1)$, 
so strictly speaking $s_1$ should also be included in the full ``one-loop" free energy results. 
This appears as a complication a priori, meaning
that at order $k$ one needs in principle the more demanding 
information from (the $\ln\mu$ coefficient of) perturbative order $k+1$. On the other hand since RG invariance is
constructed perturbatively, one may expect that the simplest minimal prescription of keeping only the
$s_k$ terms at order $\lambda^k$ should already be a good enough approximation. Accordingly, we mainly 
consider below the simplest prescription but also examine both prescriptions,
indicating the differences whenever relevant. We will see that, after the modification 
of perturbative expansion implied by RGOPT, incorporating the higher order $s_{k+1}$ at order $\lambda^k$ 
makes no crucial differences, even at one-loop order,  the resummation results being not very sensitive to such purely perturbative variations.
The same stability with respect to such variations 
was also observed at vanishing temperature in Ref. \cite{rgopt_alphas} (where
those different prescriptions were incorporated within the intrinsical theoretical
errors of the method).\\

It should be clear from the previous derivation that by construction the subtraction terms make the free energy 
perturbatively RG-invariant. But just to crosscheck it in a more pedestrian way, 
let us now reexamine the 
result at one-loop order, with the $-s_0 m^4/\lambda$ subtraction included in Eq.~(\ref{F02l}), 
using the standard one-loop RG running coupling
and mass. These are given from integrating respectively Eq.~(\ref{beta}), (\ref{gamma}) which yields 
the standard textbook result:
\be
\lambda(\mu)=\lambda(\mu_0) \left (1-b_0 \lambda(\mu_0) \ln \frac{\mu}{\mu_0} \right )^{-1} 
\simeq \lambda(\mu_0) \left (1+\frac{3}{16\pi^2} 
\lambda(\mu_0) \ln \frac{\mu}{\mu_0} +{\cal O}(\lambda^2)\right )\;,
\label{lamrun}
\ee
\be
m(\mu)=m(\mu_0) \left (1-b_0 \lambda(\mu_0) \ln \frac{\mu}{\mu_0}\right )^{-1/6} \simeq m(\mu_0) 
\left (1+\frac{1}{32\pi^2} 
\lambda(\mu_0) \ln \frac{\mu}{\mu_0}+{\cal O}(\lambda^2)\right )\;,
\ee
where $\lambda(\mu_0)$ and $m(\mu_0)$ are the coupling and mass at some arbitrary reference scale 
$\mu_0$.
Expanding Eq.(\ref{F02l}) to first order in $\lambda$ ({\it i.e.} to order $\lambda^0$) one obtains:
\bea
&(4\pi)^2 {\cal F}_0 \simeq m^4(\mu_0) \left[-\frac{8\pi^2}{\lambda(\mu_0)} 
-\frac{1}{8}( 4\ln\mu +3+\cdots) +8\pi^2 (\frac{3}{16\pi^2}\ln \mu -\frac{2}{16\pi^2} \ln \mu) +{\cal O}(\lambda)\right]
+\mbox{thermal part} \nn \\
&  \simeq m^4(\mu_0) \left[-\frac{8\pi^2}{\lambda(\mu_0)} 
-\frac{3}{8} +{\cal O}(\lambda)\right]\;,
\label{cancel0}
\eea
(where $\cdots$ stands for $\mu$-independent terms). This relation explicitly displays the cancellation, up to terms of higher order
${\cal O}(\lambda)$, of all $\ln\mu$ contributions 
coming respectively from the original one in Eq.~(\ref{F02l}), and 
from the running coupling and mass (third and fourth terms respectively in Eq.(\ref{cancel0})). 
We have neglected above the thermal contributions, but including these
does not alter the results since the exact $T$-dependent contribution does not depend explicitly on $\mu$. 
Alternatively, when the high-$T$
expansion is considered, the explicit $\ln\mu/m$ in Eq.~(\ref{F02l}) is replaced by a  $\ln\mu/(4\pi T)$ 
with the same coefficient consistently.
It is instructive to push this exercise a step further 
now incorporating in the free-energy (\ref{F02l}), restricted at one-loop order, 
all thermal contributions in the high-$T$ limit, from
Eq.(\ref{J0exp}), and plugging into the resulting expression  $m=m_D$, 
the standard perturbative thermal mass, 
Eq.(\ref{mD1}) also restricted at first order. Then, the expression 
only depends on the coupling, and using Eq.~(\ref{lamrun}) gives the result:
\be
(4\pi)^2 {\cal F}_0 \simeq T^4 \: \left[-\frac{8\pi^2}{45} +\frac{\pi^2}{72} \lambda(\mu_0) 
-\frac{\pi}{36\sqrt{6}} \lambda^{3/2}(\mu_0) +{\cal O}(\lambda^2,\ln\mu)\right ] \;,
\label{F0Trg}
\ee
where all $\mu$ dependence has been cancelled out
up to order $\lambda^2$, since $m^4_D\sim \lambda^2$. This was expected since 
the subtraction takes care of the $T=0$ lowest orders $\mu$-dependence as shown previously, while
the thermal contribution $\propto J_0(m/T)$ in Eq.~(\ref{F02l}) does not depend on the scale. More 
interestingly, one recognizes from Eq.~(\ref{F0Trg}) the standard perturbative expansion for the pressure which, upon using the 
common normalization with the free gas pressure $P_0=\pi^2 T^4/90$ and $\lambda_0=4! g^2$, can be expressed as:  
 \be
\frac{P}{P_0} = 1-\frac{15}{8} \frac{g^2}{\pi^2} +\frac{15}{2} \frac{g^3}{\pi^3} \;
\label{PP0stand}
\ee
In particular, the term $-s_0/\lambda$ is 
incorporated in the final result Eq.~(\ref{F0Trg}) once using Eq.(\ref{mD1}) since 
$-s_0 m^4_D/\lambda \sim {\cal O}(\lambda)$. The latter gives a contribution $(15/8)\,g^2$ to $P/P_0$, 
making this complete one-loop expression consistent with the standard perturbative expression of the pressure up to
order $\lambda^{3/2}$, 
while the original (unsubtracted) perturbative one-loop expression is not, giving an expansion similar 
to (\ref{PP0stand}) but with a twice too large $g^2$ term $-(15/4)\, g^2$. 
However, this agreement is merely an accident of one-loop order: at two-loop order, 
including the next order subtraction term from (\ref{sub}) does not give the correct {\em massless}
perturbative pressure when replacing $m$ by $m_D$. (In particular the subtraction $-s_0\, m^4/\lambda$ happens
to be exactly cancelled by the  hard contribution of order $\lambda$, $\propto \lambda T^4$). 
But this is not surprising, since more generally
the perturbative {\em massless} pressure~\cite{mdeb} 
cannot be obtained consistently from simply replacing the perturbative thermal mass 
(\ref{mD1}) in the expression of the {\em massive} 
pressure. 
However, when the mass is traded for a variational parameter, 
like in the OPT/SPT resummation approaches to be recalled next, one may recover the massless pressure results
under specific conditions. This is made possible
because the OPT construction and mass optimization drastically modifies the massive contributions as compared with 
the original perturbative expansion.\\

To summarize this section, starting with Eq.~(\ref{F02l})  including ${\cal E}_0$ 
from Eq. (\ref{sub}) to obtain a perturbatively RG-invariant free energy 
provides a sound basis for more elaborate resummation procedures 
like the OPT method to be addressed next. As above mentioned, an extra 
advantage of the subtraction terms (\ref{sub}) 
starting with $1/\lambda$, is that the optimization procedure will 
provide a non-trivial  mass $\t m(\lambda)$ already at the lowest one-loop order, in contrast 
with the standard SPT and HTLpt approaches (where at one-loop order the mass optimization gives a 
trivial $\lambda$-independent solution~\cite{SPT,SPT3l,htlpt1}). 
This is a welcome feature specially for more involved theories like thermal QCD 
where higher order contributions are challenging to evaluate, and a 
comparison between successive orders is crucial to establish the stability of the resummation results. 
\section{Optimized Perturbation Theory (OPT)}
The basic feature of the optimized perturbation theory (OPT) (appearing also under different names and variations~\cite{OPT-LDE,odm,pms}), 
is to introduce an extra parameter $0<\delta<1$, which interpolates between ${\cal L}_{free}$ and 
${\cal L}_{int}$ in Eq.(\ref{Lphi4}), so that the mass 
 $m$ is traded for an arbitrary trial parameter.
This is perturbatively equivalent
to taking any standard perturbative expansions in $\lambda(\mu)$, after renormalization, reexpanded  
in powers of $\delta$ after substituting:
\be m \to  m\:(1- \delta)^a,\;\; \lambda \to  \delta \:\lambda\;.
\label{subst1}
\ee
This procedure is consistent with renormalizability~\cite{gn2,chiku,optON} and gauge invariance~\cite{qcd2}, whenever the latter is relevant,
provided of course that the above redefinition of the coupling is performed
consistently for all interaction terms and counterterms
appropriate for renormalizability and gauge invariance in a given 
theory\footnote{Contrary to what is sometimes 
claimed (and worked out) in the literature, the OPT/SPT/HTLpt does not need any 
extra counterterms besides the standard ones of the corresponding massive theory: 
in particular all seemingly new divergences generated at arbitrary orders from the first replacement
in Eq.~(\ref{subst1}) are evidently related to the single standard mass counterterm. Moreover at arbitrary orders, 
temperature-dependent divergences and associated counterterms should not appear, as expected from general principles, 
provided that one keeps the mass as an arbitrary parameter and carefully subtract all nested subdivergences 
until renormalization has been completed, before using a gap-equation giving perturbatively $\t m \propto T\,\sqrt\lambda$.}.   
Note that in Eq.~(\ref{subst1}) we have introduced an extra parameter, $a$, to reflect a priori a certain freedom 
in the interpolation form. As will be demonstrated below this parameter  plays an essential role within our method for  allowing  compelling 
constraints to be imposed.  
Applying Eq.~(\ref{subst1}) to some given renormalized 
perturbative expansion for a physical quantity, $P(m,\lambda)$, reexpanding in $\delta$ 
to order $k$, and taking {\em afterwards} the $\delta\to 1$ limit (to recover the original {\em massless} theory)
leaves a remnant $m$-dependence at any finite $\delta^k$-order. 
The arbitrary mass parameter $m$ is then most conveniently fixed by a variational optimization  prescription known as the principle of minimal sensitivity  \cite{pms}
\be
\frac{\partial}{\partial\,m} P^{(k)}(m,\lambda,\delta=1)\vert_{m\equiv \tilde m} \equiv 0\;,
\label{OPT}
\ee  
thus determining a nontrivial optimized mass $\t m(\lambda)$, with nonperturbative $\lambda$-dependence, 
realizing dimensional 
transmutation (more precisely, {\it e.g.} for asymptotically free theories at vanishing 
temperatures, the optimized mass is automatically of the order of the basic scale 
$\Lambda \sim \mu\, e^{-1/(b_0\,\lambda)}$, in contrast with the original vanishing mass).  \\
In simpler ($D=1$) models, at vanishing temperatures, the procedure may be seen as a particular case of 
``order-dependent mapping''  \cite{odm},  
which has been proven \cite{deltaconv} to converge exponentially fast for the $D=1$ $\phi^4$ oscillator energy levels.  
For higher dimensional $D > 1$ renormalizable models, no rigorous convergence proof exists, although the OPT was shown to partially  
damp the factorially divergent (infrared renormalons) perturbative behavior at large orders~\cite{Bconv}.   
Nevertheless, this technique  can give rather successful  approximations to some nonperturbative quantities beyond the large-$N$ (or mean field) approximations in a large variety of  physical situations which 
include the study of phase transitions within condensed matter related renormalizable models\cite{optGN,tulio,beccrit,bec2}  
as well as within QCD non renormalizable effective models  \cite{njlprc}.\\
We emphasize  that at finite temperatures  the very same basic idea has been exploited 
by the SPT~\cite{SPTearly,SPT}/HTLpt~\cite{htlpt1} method, where in this thermal 
context the screening thermal mass is treated as an arbitrary variational parameter,
and in Eq.~(\ref{OPT}) $P$ also depends on $T$ like e.g., Eq.~(\ref{F02l}). 
\section{Renormalization Group compatibility of OPT} 
In most previous standard OPT (or similarly SPT and HTLpt) applications, 
the so-called linear $\delta$-expansion is used, assuming $a=1/2$ (i.e. $m^2\to m^2 (1-\delta)$
for a scalar mass) 
in Eq.(\ref{subst1}) mainly for simplicity and economy of parameters while
the more recent approach, developed in Refs. \cite{rgopt1,rgopt_Lam,rgopt_alphas}, 
differs in two crucial aspects which turn out to drastically improve the convergence.
First, it introduces a straightforward marriage between  OPT and renormalization group (RG)  
properties, by  requiring the ($\delta$-modified) expansion to satisfy, in addition to the OPT Eq.(\ref{OPT}), a 
standard RG equation:
\be
\mu\frac{d}{d\,\mu} P^{(k)}(m,\lambda,\delta=1)=0, 
\label{RG}
\ee 
where the RG operator was defined in Eq.~(\ref{RG}).
Moreover, once combined with Eq.(\ref{OPT}), the RG equation takes the reduced massless form:
\be
\left[\mu\frac{\partial}{\partial\mu}
+\beta(\lambda)\frac{\partial}{\partial \lambda}\right]P^{(k)}(m,\lambda,\delta=1)=0\;.
\label{RGred}
\ee
Therefore, Eqs.~(\ref{RGred}) and (\ref{OPT}) if used together, completely determine
{\em optimized} $m\equiv \t m$ and $g\equiv \t g$ ``variational'' fixed point values.\\
Since interaction and free terms from the original perturbative series are rather drastically 
reshuffled by the modification implied by Eq.~(\ref{subst1}), the RG invariance is in general 
no longer perturbatively satisfied, even when the original perturbative 
series is RG-invariant prior to performing (\ref{subst1}). 
This spoiled RG invariance has to be restored in some manner, and thus
Eq.~(\ref{RG}) gives a nontrivial additional constraint. 
This feature has been seldom  appreciated and considered in many 
former applications of the $\delta$-expansion/OPT method to renormalizable theories 
(perhaps in part because in many analyses with more elaborated theories the OPT is 
restricted to first order, where RG improvements are supposed to play a minor role). 
This important role of RG properties was recognized much earlier in Refs. \cite{qcd1,gn2} where to recover the RG consistency  
 the standard linear $\delta$-expansion was resummed to all orders. Indeed, this resummation 
can be done, at least for the pure RG dependence up to two-loop, but the result 
comes as a rather involved integral 
representation, not practically intuitive and making
difficult to perform the mass optimization or  
to generalize to other physical quantities and other models. In contrast, the 
purely perturbative procedure together with Eq.(\ref{RG}) appears  a 
considerable shortcut, straightforward to apply to any model. Intuitively, 
just as the stationary point solutions from Eq.~(\ref{OPT}) are expected to give sensible approximations, 
at successive orders, to the actually massless theory, one similarly expects that combining the latter
with the RG solutions should further give a sensible sequence of best approximations to the exactly scale
invariant all order results. \\
Still, a well-known drawback of the standard OPT approach is that, beyond lowest order, solving 
Eq.~(\ref{OPT}) generally gives more and more solutions 
at increasing orders, some of which are likely to be complex-valued. 
More generally, without some insight on the nonperturbative behavior of the solutions, it may be difficult
to select the right one, and the unphysical complex-valued optimized solutions at higher orders are embarrassing. 
This is incidentally a problem encountered first at three-loop order in SPT \cite{SPT3l} 
and HTLpt applications to QCD\cite{HTLPT3loop}. The mass optimization is then replaced
by alternative prescriptions, most often using simply the purely perturbative screening mass, 
but accordingly loosing a more nonperturbative ingredient from the optimized mass. 
But RG considerations also provide a possible way out, which 
is the second main difference and new feature of the present RGOPT version.  
For QCD a compelling selection criterion was proposed, in which only the solution(s) 
{\em continuously matching} the standard perturbative (asymptotically free) 
RG behavior for vanishing coupling are retained~\cite{rgopt_Lam,rgopt_alphas}.
This prescription can easily be generalized to any model, like the non asymptotically free (AF) $\phi^4$ theory, 
by similarly requiring to asymptotically match the solutions to the standard perturbative behavior for small coupling, 
namely for fixed $\t m$ and arbitrary scale $\mu$:
\be
\t \lambda (\mu \ll \t m) \sim \left (b_0 \ln \frac{\t m}{\mu}\right )^{-1} +
{\cal O}\left ( \ln^{-2} \left (\frac{\t m }{\mu}\right )\right)\;.
\label{rgasympt}
\ee
At zero temperature this turns out to give a unique solution for both the RG and OPT equations, up to rather high orders. 
An additional welcome feature is that by
requiring at least one RG solution to fulfill Eq.~(\ref{rgasympt}) leads to a strong necessary condition on the basic interpolation, 
Eq.~(\ref{subst1}), uniquely determining $a$ from the universal (scheme-independent) first order RG coefficients: 
$ a\equiv \ga_0/b_0$, as we derive in more detail below.
A connection of the OPT exponent $a$ with RG anomalous dimensions/critical exponents  had also been established in a very different context, in the  $D=3$ $\phi^4$ model for the Bose-Einstein condensate (BEC) 
critical temperature shift by two independent OPT approaches~\cite{beccrit,bec2}, where it also led to real OPT solutions~\cite{bec2}.  
However,  AF-compatibility and reality of solutions can appear to be  mutually exclusive beyond lowest order, depending on the particular model. 
A simple way out is to further exploit the RG freedom, considering a perturbative renormalization scheme change to attempt to 
recover  RGOPT solutions both AF-compatible and real~\cite{rgopt_alphas}. We will see that this extra complication is not even necessary
in the present case, where (at least up to the two-loop order) the RG-compatible solutions remain real for a large range of relevant
values of the coupling and temperature. 
All these features are easy to generalize at finite temperatures due to the fact that RG properties are essentially determined by the
divergence structure of the $T=0$ part. So, the only complication is technical since at finite temperature the previous Eqs.~(\ref{OPT}), (\ref{RG}), 
and (\ref{RGred}) come with an extra $T$ dependence. Let us now illustrate explicitly all those features by evaluating the RGOPT modification of
the free energy of a thermal scalar field. 
\section{1-loop, ${\cal O}(\delta^0)$}
\subsection{$T=0$}
Let us first truncate Eq.~(\ref{F02l}) at strict one-loop order, 
and first restricting to $T=0$ which is simpler and sufficient
to determine the RG-exponent $a$ in Eq.(\ref{subst1}). We have
\be
(4\pi)^2 {\cal F}^{\rm RGI}_0(T=0) = -\frac{s_0}{\lambda} m^4 -\frac{1}{8} m^4 \left (3+2\ln\frac{\mu^2}{m^2} \right ) \;,
\ee
where  the superscript RGI emphasizes the (perturbative) RG invariance of this quantity. 
At this order the calculation is elementary so it can best illustrate the main steps. 
Now, applying Eq. (\ref{subst1}), performing the $\delta$-expansion to order $\delta^0$ consistently, and taking {\em afterwards} $\delta\to 1$,
gives:
\be
(4\pi)^2 {\cal F}^{\rm RGI}_0(\delta^0,\delta=1) = 
m^4 \left[-\frac{s_0}{\lambda}(1-4a) -\frac{1}{8} \left (3+2\ln\frac{\mu^2}{m^2} \right )\right]  \;.
\label{F0opt0}
\ee
Note that the ${\cal O}(1)$ term remained unmodified (this is a general property of OPT: 
expanding at order $\delta^k$ and taking $\delta\to 1$ leaves the order $\lambda^k$ term
unaffected due to the screening from $\lambda\to \delta \lambda$). Then, requiring 
Eq.~(\ref{F0opt0}) to be perturbatively RG-invariant after this modification of the perturbative series, 
i.e. applying the RG Eq.~(\ref{RGred}), on gets
\be
m^4 \left[ \left (1-\frac{b_0}{\gamma_0} \,a\right ) +{\cal O}(\lambda)\right] = 0\;,
\label{RG0}
\ee
which uniquely fixes 
\be
a=\frac{\gamma_0}{b_0} =\frac{1}{6} \;,
\label{acrit}
\ee
where, in Eqs. (\ref{RG0}) and (\ref{acrit}), we made the RG coefficient dependence explicit to emphasize the generality of these results.
At this point several remarks are in order: 
\begin{itemize}
 \item 1) the very same result, Eq.~(\ref{acrit}), was obtained~\cite{rgopt_Lam,rgopt_alphas}  (up 
to a trivial $b_0$ difference of normalization by a factor 2), while considering the RGOPT for QCD  (with appropriate QCD values for those RG coefficients). 
The exponent $a$ is universal for a given model, in the sense that it only depends on 
the first-order RG coefficients, which are renormalization scheme independent. Furthermore, at vanishing temperature, Eq.~(\ref{acrit})
greatly improves the convergence of the procedure at higher orders: considering 
only the first RG coefficients $b_0, \gamma_0$ dependence (i.e., neglecting higher RG orders and non-RG terms), 
it gives the exact nonperturbatively resummed result at the first $\delta$ order and any
successive order~\cite{rgopt_alphas}. This is not the case for $a=1/2$ (for a scalar mass), where the convergence appears
very slow, if any. 
\item 2) The standard linear $\delta$-expansion interpolation, widely used for zero temperature models
and for SPT/HTLpt, takes $a=1/2$ \cite{OPT-LDE,optON,SPT3l,OPT3l}:
\be
m^2 \to m^2 (1-\delta)\;\;,
\label{LDE}
\ee
thus our RG-compatible exponent (\ref{acrit}) is three times smaller~\footnote{In QCD e.g for 3 light quark flavors $a=\gamma_0/(2b_0)=4/9$
is also substantially smaller than the linear case value $a=1$ for fermion masses.}.
Indeed the standard OPT/SPT interpolation (\ref{LDE}) leads to an unmatched RG equation, while the OPT equation (\ref{OPT}) solved
for $\lambda(m)$ (or equivalently for $m(\lambda)$ gives, in the perturbative regime $\mu \ll m$:
\be
\lambda(\mu \ll m) \sim -\frac{16\pi^2}{\ln (\frac{m}{\mu})}\;,
\ee
in clear contradiction with the true running by a wrong overall sign plus a factor three too large.
\item 3) At the very first non-trivial $\delta^0$ order, once having fixed $a=\gamma_0/b_0$ the RG equation is satisfied and thus does not give further constraint.
We will see that at the next  and higher orders in $\delta$, Eq.~(\ref{acrit}) is always required for the RG equation to have 
at least one solution matching Eq.~(\ref{rgasympt}). In addition, it also fixes $\lambda$ in terms
of the other parameters ($m$, and the only remaining parameter $\mu/T$ when considering the thermal part). 
\end{itemize}
Let us next consider the other OPT constraint given by  Eq.~(\ref{OPT}). 
Still neglecting the thermal part, and pulling out an overall factor yields
\be
m^3 \left[ \frac{1}{b_0\,\lambda} +\frac{1}{2}\left (1+\ln \frac{\mu^2}{m^2} \right) \right] = 0\;.
\label{opt0}
\ee
One readily remarks the explicit exact scale-invariance of Eq.(\ref{opt0}), thus of its solution, 
provided that one uses for $\lambda\equiv\lambda(\mu)$ 
the exact (one-loop resummed) running in  Eq.~(\ref{lamrun}), since the expression
$1/\lambda(\mu)+b_0 \ln \mu $ is explicitly
$\mu$-independent. 
Letting apart the trivial $m=0$ solution, Eq.(\ref{opt0}) has the more interesting solution
\be
\t m^2=\mu^2 e^{1+\frac{2}{b_0\lambda}}\;\;,
\label{solopt0}
\ee
which is seen to be compatible, for $\lambda\to 0$, 
with the perturbative first order standard RG solution
obtained from solving  Eq.(\ref{beta}) for $\lambda(\mu)$ at first order. Namely,
for fixed $m$ and arbitrarily small $\mu$, it exhibits infrared freedom: $\lambda(\mu\ll m)\simeq
(b_0 \ln (m/\mu))^{-1}$. \\
Moreover, plugging Eq.~(\ref{solopt0}) within 
the modified vacuum energy expression (\ref{F0opt0}) for the case of vanishing temperatures, gives
a remarkably simple result:
\be
{\cal F}_0(\tilde m,\lambda) = -\frac{\tilde m^4}{8\:(4\pi)^2 }\;.
\label{F0optT0}
\ee
Despite its somewhat trivial look, Eq. (\ref{F0optT0}) represents a nonperturbative result, in the sense that it 
only involves the expression of $\tilde m^4$ from Eq.~(\ref{solopt0}). 
Apart from the one in Eq.~(\ref{solopt0})
all $\lambda$-dependence disappeared (in particular
the $-s_0/\lambda$ term has consistently disappeared upon using the OPT gap equation (\ref{opt0})). 
Therefore, Eq.~(\ref{F0optT0}) gives a non-trivial $T=0$ (negative) vacuum energy contribution
and resembles much the large-$N$ nonperturbative result, up to 
appropriate identification of $b_0\lambda$, but here obtained from the OPT. When higher RG orders and non-RG
contributions are included, they spoil this simple form result~\cite{rgopt_alphas}, as does also the thermal part, 
that we will consider next. In these cases there are remnant coupling dependences, 
once having used the solution of Eq.~(\ref{OPT}) within the physical expression 
of the vacuum energy. 
\subsection{$T\ne 0$}
Let us now consider the thermal contributions in Eq.~(\ref{F02l}) still at one-loop order. After performing the $\delta$-expansion
to the lowest corresponding order $\delta^0$ we obtain the $T\ne0$ free energy, similarly as in (\ref{F0opt0}):
\be
(4\pi)^2 {\cal F}^{\rm RGI}_0(T\ne 0,\delta^0,\delta=1) = 
m^4 \left[-\frac{1}{2b_0\,\lambda} -\frac{1}{8} \left (3+2\ln\frac{\mu^2}{m^2} \right )\right] -\frac{1}{2}T^4 J_0\left(\frac{m}{T} \right) \;.
\label{F0optT}
\ee
where the first term $-s_0 m^4 (1-4a)/\lambda =-m^4/(2b_0\,\lambda)$ is the only one affected by (\ref{subst1}) 
at this lowest order.  
Calculations are slightly more involved than for $T=0$ but 
note that the $\delta$-expansion and subsequent 
OPT minimization equation involve successive derivatives of the thermal function $J_n(m/T)$. 
For more generality and to make contact with various other resummation methods, it turns out to be
particularly convenient to express all our RGOPT results 
in terms of the one-loop renormalized self-energy, including all thermal
dependence:
\be
\Sigma_R \equiv \frac{\lambda}{2} \int_p \frac{1}{p^2+m^2} +\Sigma^{\rm ct}
=\gamma_0 \lambda \left[m^2 \left (\ln\frac{m^2}{\mu^2} -1\right )+T^2 J_1\left (\frac{m}{T}\right )\right]\;,
\label{Sigma}
\ee
(where for completeness the mass counterterm reads $\Sigma^{\rm ct}$ reads 
$\gamma_0 \lambda m^2/(2\epsilon)$). 
This simple factorization is possible for the scalar $\phi^4$ model up to the two-loop level, because the 
two-loop contribution (the last order $\lambda$ term in Eq.~(\ref{F02l})) factorizes 
as the square of one-loop expressions (i.e. graphs with a different ``nested" topology only appear at the three-loop level 
for the $\lambda \phi^4$ interactions). 

Then, noting that $\frac{\partial}{\partial m^2} \int_p \ln (p^2+m^2)=2 \Sigma_R/\lambda$, the 
exact solution of the OPT Eq.~(\ref{OPT}) can easily be written in the form of a self-consistent ``gap"
equation for $\t m$:
\be
\t m^2 = (4\pi)^2\,b_0\:\Sigma_R = b_0 \frac{\lambda}{2} \left[ \t m^2 
\left (\ln  \frac{\t m^2}{\mu^2} -1\right ) +T^2 J_1\left (\frac{\t m}{T}\right ) \right] \;\;, 
\label{opt0Tex}
\ee
which like the $T=0$ previous case, is exactly scale-invariant by construction, 
as we will illustrate more explicitly below.
\subsubsection{Digression: connection with large-$N$ and 2PI results}
As an important digression, we point out that Eq.~(\ref{opt0Tex})  is recognized as the very same solution
obtained for the large-$N$ $O(N)$ $\phi^4$ model in Ref.~\cite{phi4N}, upon appropriate 
$b_0$ definition for the large-$N$ case. Indeed, in the leading $1/N$
approximation, the only contributing graphs have the one-loop
structure, and the mass-gap equation can be solved exactly. More precisely it is easily found from 
the arbitrary $N$ RG coefficients given {\it e.g.} in Ref.~\cite{RGphi4loop} that in our normalization, 
\be
(4\pi)^2\,b_0(N)=\frac{N+8}{3},\;\;(4\pi)^2\,\gamma_0(N)=\frac{N+2}{6} , 
\ee
so that in particular we have for the crucial exponent in (\ref{subst1}):
\be
a\equiv \frac{\gamma_0}{b_0} = \frac{1}{2} \left(\frac{N+2}{N+8} \right) \to \frac{1}{2}\; \mbox{for}\; N\to\infty
\ee
for which values all our previous construction, and the corresponding mass gap equation in (\ref{opt0Tex}),
reproduce exactly the large $N$ results 
in \cite{phi4N}. Note in particular that $a=1/6$ for $N=1$ while accidentally the large $N$ value of $a=1/2$ is the standard linear one,
but here being fully consistent with RG properties. The fact that the one-loop RGOPT reproduces exactly the
large $N$ result can be seen as the finite temperature analog of similar RGOPT properties~\cite{rgopt1} obtained 
for the large $N$ limit of the GN model. \\
Similarly, Eq.~(\ref{opt0Tex})  is also recognized as the very same form of mass gap solution 
obtained in the 2PI formalism but at two-loop order\cite{2PI} (or also in the tadpole approximation 
for the self-energy~\cite{2PItadpole}), except that in \cite{2PI} the correct 
$b_0=3/(16\pi^2)$ is effectively replaced by $b_0/3$ because, as explained by the authors, only one channel out of three
is taken into account at this level of the 2PI approximation, similarly to the leading $1/N$ 
approximation. We will come back below on this apparent $b_0$ value mismatch
when discussing the perturbative reexpansion of the pressure to make contact with standard perturbation
results. In fact, the analogy
with Ref.~\cite{2PI} goes further, in particular their expression of the mass gap solution
is also exactly scale invariant at two-loop order, and the free energy involves,  
after renormalization, a term $-m^4/(2\lambda)$, once again identical to our subtraction
terms $-m^4/(32\pi^2 b_0\lambda)$ in Eq.~(\ref{F0opt0}), 
when taking $b_0\to b_0/3$. But the origin of this $1/\lambda$ term in ref.~\cite{2PI} is
very different, emerging upon using the gap equation. 
Incidentally, the scale-invariance of the 2PI results
 is essentially due to a renormalization procedure that is peculiar to the scalar model, 
 inspired by the large $N$ limit~\cite{phi4N}, which thus appears to us as being accidental, 
 and limited to the first non-trivial two-loop level. 
 In contrast the mass gap (\ref{opt0Tex}) is obtained already at the one-loop level, 
 and it should be clear from the previous construction
 that the RGOPT systematic 
 procedure is applicable in any model at arbitrary orders.
\subsubsection{One-loop RGOPT mass gap and pressure solutions}
Although Eq.~(\ref{opt0Tex}) may easily be solved numerically, it is 
instructive to consider next the approximation
given by the high-$T$ expansion, which is very precise as long as $T \gsim m$: in fact this condition
can be easily checked a posteriori considering the optimized solution $\t m$. At one-loop level, it turns 
out that the optimized mass always satisfies this criterion for all the relevant range
of coupling values. Therefore, we will not need to solve
the exact Eq.~(\ref{opt0Tex}) for all practical purposes. 
In the high-$T$ approximation (\ref{J0exp}), the one-loop order OPT Eq.~(\ref{OPT}) produces 
(discarding the trivial solution $\t m=0$) a simple quadratic equation for $m$:
\be
 \left (\frac{1}{b_0\,\lambda} +L_T \right )\, x^2 +2\pi\; x -2\frac{\pi^2}{3} =0\;,
\label{opt0T} 
\ee
where  $m \equiv x T$ and we defined for shorthand notations the $\mu/T$ dependent part 
$L_T\equiv \ln [\mu/(4\pi T)e^{\gamma_E}]$. As already explained above the RG Eq.~(\ref{RGred})  reduces to Eq.~(\ref{RG0}) 
which is already satisfied for Eq.~(\ref{acrit}), thus it gives no additional constraint.\\ 
Solving  Eq.~(\ref{opt0T}) gives two real solutions, but one is clearly unphysical, giving
$\t m <0$ for any $\lambda$. The other unique physical solution is thus
\be
\ds \frac{\t m^{(1)}}{T} = \pi  \frac{\sqrt{1+\frac{2}{3}\left (\frac{1}{b_0\lambda}+ L_T\right )}-1}
{\frac{1}{b_0\lambda}+ L_T} \;,
\label{solopt0T}
\ee
where for more generality we kept the dependence on $b_0$ explicit. Despite the apparently minor
modification of the series represented by Eq.~(\ref{F0opt0}) at this first order, 
the solution given by Eq.~(\ref{solopt0T}) has clearly a nonperturbative dependence on $\lambda$.
We stress that the variational mass (\ref{solopt0T}) is 
unrelated to the physical screening mass~\cite{mdeb} in Eq.~(\ref{mD1}), and thus has no reason to reproduce the latter. 
Moreover, as anticipated, Eq.~(\ref{solopt0T}) is strictly {\em exactly} scale-invariant,
provided that one uses for $\lambda\equiv\lambda(\mu)$ the exact (one-loop resummed) running in Eq.~(\ref{lamrun})  
(now with $\mu$ being temperature-dependent as usual), since the expression
$1/\lambda(\mu)+b_0 L_T \equiv 1/\lambda(\mu)+b_0 \ln \mu+\cdots$ is explicitly
$\mu$-independent. In other words the mass gap in (\ref{solopt0T}) actually only depends on the single 
parameter $b_0\lambda(\mu_0)$, where $\mu_0$ is some reference scale, typically $\mu_0=2\pi T$.\\

\begin{figure}[htb]
\epsfig{figure=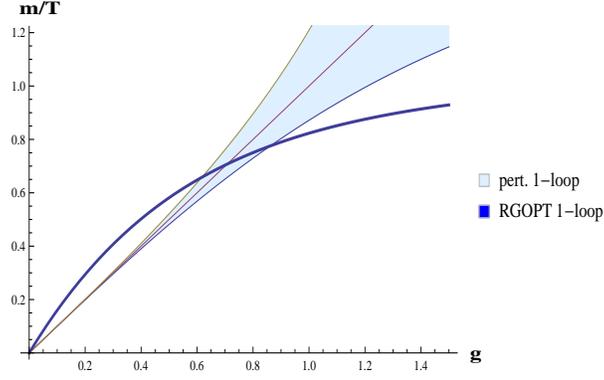,width=8cm,height=5cm}
\caption[long]{RGOPT mass $m/T$ at one-loop ($\delta^0$) order (thick line), versus standard one-loop perturbative  
mass as functions of $g(\mu_0=2\pi T) =(\lambda(\mu_0)/24)^{1/2}$ with scale-dependence. 
Grey (light blue) bands:  variation of the standard perturbative mass 
between $\mu=\pi T$ and $4\pi T$ using  the exact one-loop
running coupling in Eq.~(\ref{lamrun}).
NB: the RGOPT mass has actually zero thickness since it is exactly scale-invariant.}
\label{Mopt0T}
\end{figure}

\begin{figure}[htb]
\epsfig{figure=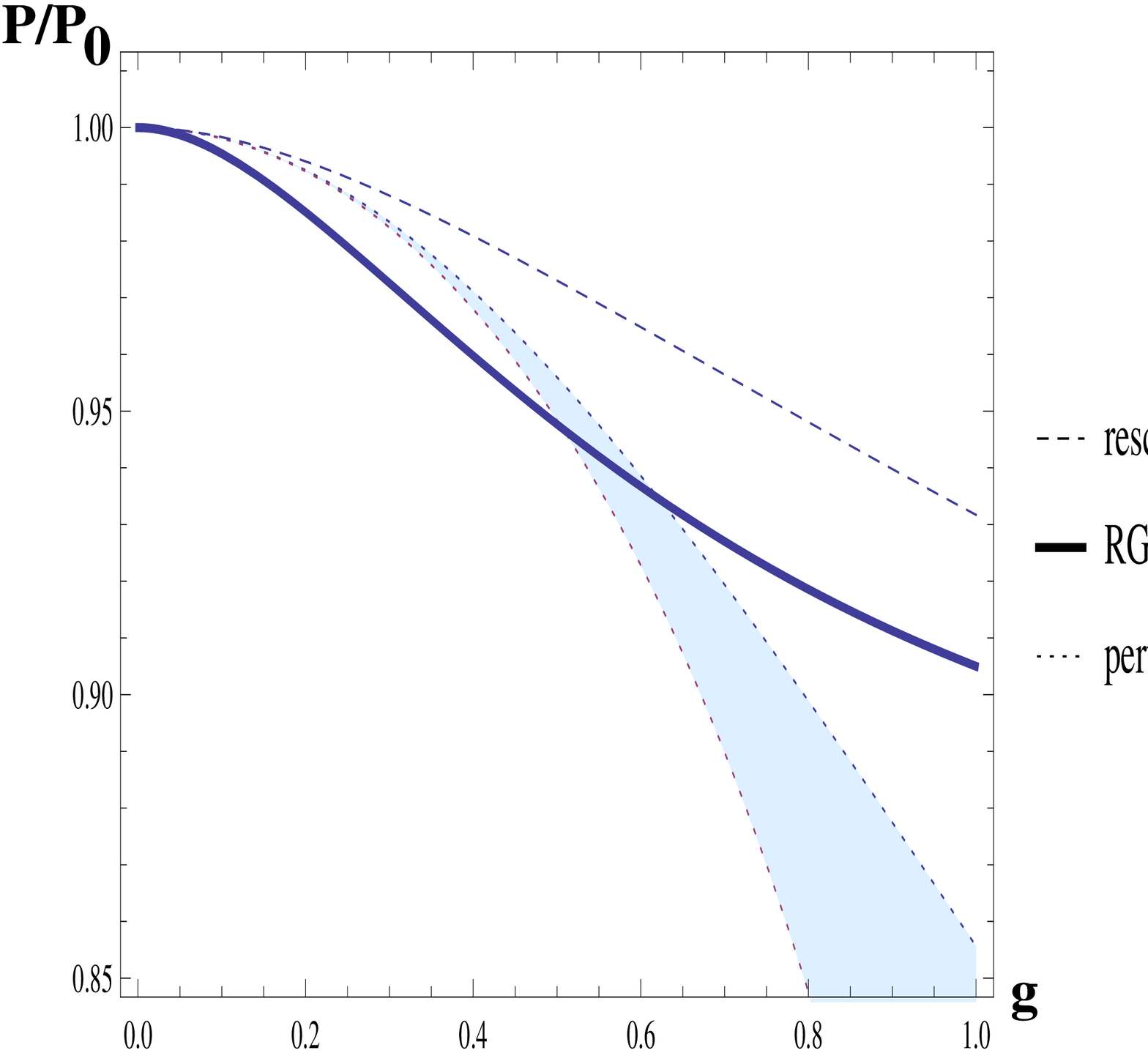,width=8cm,height=5cm}
\caption[long]{One-loop ($\delta^0$) RGOPT pressure (thick), and rescaling $\lambda\to\lambda/3$ (dashed), 
versus standard one-loop (dotted, light blue bands) pressure  
as function of $g=\sqrt{\lambda(\mu)/24}$ with 
scale-dependence between $\mu=\pi T$ and $\mu=4\pi T$. 
NB: the RGOPT pressure has actually zero thickness since it is exactly scale-invariant.}
\label{Popt0T}
\end{figure}

Before we proceed, it is worth to comment a little more on this
result: recall that prior to the $\delta$-expansion, 
the basic one-loop expression (first line in Eq.(\ref{F02l})) 
with the first term $s_0$ in the subtraction ${\cal E}_0$, 
is by construction RG-invariant to one-loop order,  
{\it i.e.} up to neglected higher order terms ${\cal O}(\lambda)$. But 
what is more remarkable is that the optimal mass resulting from solving (\ref{OPT}) 
is {\em exactly} scale-invariant to all orders (of course ``all orders" but 
neglecting genuine higher orders in the running coupling, 
i.e. keeping only the $b_0$ dependence to all orders). This is a direct consequence of the value $a=\gamma_0/b_0$ in
the interpolating relation, Eq.(\ref{subst1}). This result is the finite temperature analog  of what
was similarly obtained generically
for zero temperature QCD in Ref.~\cite{rgopt_alphas}: namely, 
``all-order'' (one-loop RG) resummed results are correctly 
obtained by the very first RGOPT $\delta$ order. Indeed, since this is a generic result, 
we anticipate that applying the same procedure to thermal QCD will give similar one-loop results,
with an OPT equation and solution very similar to Eqs.~(\ref{opt0T}) and (\ref{solopt0T}) 
up to obvious changes in some factors,
but also exhibiting exact scale-invariance). 
However, this exact 
scale-invariance is due to the form of the exact one-loop running of the coupling, perfectly matching Eqs.~(\ref{opt0T}), 
(\ref{solopt0T}), which does not generalize once including higher RG orders in the $\beta$ function
and non-RG dependence at higher orders. As we examine in next section, at the two-loop ${\cal O}(\lambda)$ order, 
the scale invariance resulting from RGOPT extends beyond the two-loop perturbative order 
at which it is imposed by construction, but a (moderate) scale dependence reappears unavoidably 
at a {\em finite} higher perturbative order, precisely at order $\lambda^3$, thus one order higher than naively expected. \\
To proceed one may expand Eq.~(\ref{solopt0T}) 
perturbatively, which is easily seen to be an expansion
in $\sqrt\lambda$, as expected. One then finds:
\be
\frac{ { \t m}^{(1)} }{T} \sim \pi \left[ \sqrt{\frac{2}{3}}\sqrt{b_0 \lambda} -b_0 \lambda 
+\frac{1}{2\sqrt 6}(3-2L_T) (b_0 \lambda)^{3/2}+ L_T\,(b_0\lambda)^2 +\cdots \right]\;,
\label{solopt0Tpert}
\ee
where we kept the $b_0$ dependence explicit on purpose. \\
 As an important side remark, up to now we have considered the simplest minimal prescription of incorporating 
 only the $-m^4\,(s_0/\lambda)$
subtraction at one-loop order, strictly necessary for recovering perturbative RG invariance. 
It is thus opportune to mention what is changing if incorporating the next order subtraction $s_1\ne 0$
term from (\ref{sub}), being formally also of one-loop order. In fact this simply amounts to the replacement  
$L_T\to L_T+2s_1=L_T-2$ consistently in all previous expressions (\ref{opt0T}), (\ref{solopt0T}), and (\ref{solopt0Tpert}), as could be easily traced by
consistently introducing $-m^4\,s_1$ into Eqs.~(\ref{F0optT}), (\ref{opt0Tex}).
Therefore, it means that at one-loop RGOPT order $s_1$ can be simply absorbed by a change of scale 
(or renormalization scheme) definition, 
$\mu \to \mu\, e^{2s_1}= \mu\, e^{-2}$. Thus apart from changing the reference scale with respect to the $\ms$-scheme, 
it does not really change physical results: if redefining accordingly the coupling with a RG evolution $\mu \to \mu\, e^{-2}$, 
we obtain strictly identical results.
With this in mind, for now on we proceed with the simplest choice $s_1=0$  at one-loop.\\
Coming back to Eq.(\ref{solopt0T}) it is obviously 
an expansion solely in the single parameter $(b_0\lambda)$. Using the $b_0$ value 
from Eq.~(\ref{rgcoeff}) we have:
$\pi \sqrt{2b_0/3}=1/(2\sqrt 2)$, which tells that   
the first order coefficient differs from the standard Debye 
mass in (\ref{mD1}), $m^2_D\sim (\lambda/24) T^2$, being $\sqrt 3$ larger. This originates directly from
the correct value $b_0 =3/(16\pi^2)$ used in $a=\gamma_0/b_0$ in (\ref{subst1}), which is the only value compatible with RG invariance. 
The factor 3 in $b_0$ is the statistical factor
originating from three similar graphs contributing to the $\beta$ function, as is well-known. 
Thus, the first perturbative coefficient of the Debye screening mass would be
obtained from Eq.~(\ref{solopt0T})
if one would take $b_0/3=1/(16\pi^2)$ as given by a single loop
contributing to the self-energy at one-loop, as argued in Ref.~\cite{2PI}. 
The standard perturbative term of order $ \lambda$, comparing with Eq.~(\ref{mD1}), is also reproduced provided again
that one takes $b_0=1/(16\pi^2)$ in Eq.~(\ref{solopt0T}). But, as mentioned above,
Eq.~(\ref{solopt0T})-(\ref{solopt0Tpert}) reproduce exactly (at arbitrary orders) the large 
$N$-results (e.g. Eq.~(5.7) of ref.~\cite{phi4N}), as can be checked upon identifying the correct
large-$N$ value of $b_0 =1/(16\pi^2)$ in the normalization of \cite{phi4N}.
This factor $3$ discrepancy in the optimized mass  for the $N=1$ $\phi^4$ model
from $b_0$ mismatch is not a problem, since the OPT nonperturbative variational mass, not being a physical parameter, 
has no physical connection with the perturbative physical screening 
mass in Eq.~(\ref{mD1}) and is therefore not required to reproduce the latter. 
Indeed, there are $\ln (\lambda)$ terms appearing at the three-loop order
in the genuine screening mass\cite{Trev,SPT3l}, Eq.~(\ref{mD1}), 
that are not present in the expansion of Eq.~(\ref{solopt0T})
which only involves $\lambda$ and $\lambda^{1/2}$ powers. Incidentally 
the fact that the standard OPT or SPT/HTLpt correctly reproduces the first two orders of the thermal 
perturbative mass expansion~\cite{SPT,htlpt1} appears in retrospect merely 
accidental, due to the common canonical choice $a=1/2$ in (\ref{subst1}), i.e. {\em as if} one 
had taken $a=\gamma_0/b_0$ with $b_0\to b_0/3$.\footnote{We anticipate that
for QCD, it happens accidentally that $\gamma_0^G(\rm QCD)=b_0(\rm QCD)/2$, where $\gamma_0^G(\rm QCD)$ is the gluon anomalous
mass dimension easily calculable from the relevant counterterm given e.g. in \cite{htlpt1,HTLPT3loop}. Thus $a=1/2$ for the gluonic contributions 
so that the analoguous RG-compatible OPT mass (\ref{solopt0Tpert}) will coincide for the first few perturbative
terms with the QCD gluon screening mass~\cite{Trev}.}\\
 The (exact) expression given by Eq.~(\ref{solopt0T}) is plotted as a function of the coupling in 
Fig. \ref{Mopt0T} in the very common normalization\cite{Trev} 
$\lambda\equiv 24 g^2$, where it is compared to the standard purely perturbative thermal screening
mass, $m_D/T$, with scale dependence illustrations. In particular, we remark
the saturation of the optimized mass for sufficiently large coupling, 
which agrees qualitatively well with what is obtained at the two-loop order in \cite{2PI}. 
This saturation can be seen more explicitly by expanding Eq.~(\ref{solopt0T}) for strong
coupling:
\be
\frac{ { \t m}^{(1)} }{T} \sim \pi \frac{\sqrt{1+\frac{2}{3} L_T}-1}{L_T}+{\cal O}(\lambda^{-1})\;,
\ee
The above relation reveals that even if we do not expect our approximation to be valid for arbitrarily large coupling the relation 
 $\t m/T \lsim 1$ is always valid while $\t m/T \ll 1$ in the more perturbative range (see 
Fig.\ref{Mopt0T}). Therefore, the high- $T$ approximation
used to derive those analytic expressions is fully justified a posteriori. 
Concerning the scale dependence, in order to compare with standard results, 
we use the physical OPT solution, Eq.~(\ref{solopt0T}),
 replacing $\lambda\equiv\lambda(\mu)$ by its 
``exact" one-loop running coupling in Eq.~(\ref{lamrun}). We take  
 $\mu_0\equiv 2\pi T$ as a reference scale, and vary as usual the scale
$\mu$ in the range $[\pi T,4\pi T]$. 
The plot in Fig. \ref{Mopt0T} 
is only made for illustration and comparison
with the standard perturbative thermal mass, since as explained earlier  
Eq.~(\ref{solopt0T}) is  {\em exactly} scale-invariant, i.e. the RGOPT mass gap in 
Fig. \ref{Mopt0T} has actually zero thickness, being valid for {\em any} scale.~\footnote{Except
for the fact that at some large scale $\mu$, depending on $\lambda(2\pi T)$ value, one hits 
the naive (one-loop) Landau pole, more precisely at $\mu/(2\pi T)=e^{1/(b_0\lambda(2\pi T))}$: 
for instance for $g=\sqrt{\lambda(2\pi T)/24}=1$ the Landau pole is reached at $\mu/(2\pi T)\simeq 8.96$.} 
Even if one would use the approximate one-loop expanded running coupling 
(i.e. the last expression in the RHS of Eq.~(\ref{lamrun})), 
the scale dependence would be extremely moderate, barely visible on the same plot. \\
Next, coming to the pressure it has a rather simple expression at this one-loop order,
in terms of the OPT mass $\t  m$ and normalized to the ideal gas pressure $P_0=\pi^2 T^4/90$ (still keeping the general $b_0$-dependence):
\be
\frac{P^{(1)}}{P_0} = 1 -\frac{15}{4\pi^2}\frac{\t  m^2}{T^2}+\frac{15}{2\pi^3}\frac{\t  m^3}{T^3}
+ \frac{45}{16\pi^4} \left (\frac{1}{b_0\lambda}+L_T\right ) \:\frac{\t  m^4}{T^4} +{\cal O}(\t m^6/T^6)\;
\label{P1P0}
\ee
where we remind that this is actually an approximation according to using (\ref{J0exp}), which we argue is however precise at the $10^{-3}$
level up to $x =m/T \lsim 1$.
One can thus
plug the OPT mass expression, Eq.~(\ref{solopt0T}), into Eq.~(\ref{P1P0}) 
to obtain the full $\lambda$-dependence. After some algebra it takes a compact form:
\be
\frac{P^{(1)}}{P_0}(G) = 1 -\frac{5}{4}G -\frac{15}{2} G^2\,(1+G) 
+\frac{5}{3} \sqrt{6} \left[ G\left (1+\frac{3}{2}G\right )\right]^{3/2} +{\cal O}(x^6)\;,
\label{P1P0G}
\ee
 where we defined $1/G\equiv 1/(b_0\lambda(\mu))+L_T =1/(b_0\lambda(\mu_0))+\gamma_E-\ln 2$, 
to emphasize that $P$ is exactly scale-invariant and only depends on the single
parameter $\lambda(\mu_0)$. Expression (\ref{P1P0G}) also explicitly separates the
``perturbative'' first three terms from the clearly ``non-perturbative'' last term, and is valid implicitly in the high-$T$ approximation as indicated, but very precise 
as long as $x\lsim 1$, corresponding to $G\lsim 3.3$ meaning very strong coupling for $\lambda(\mu_0)\sim G/b_0$. The first neglected term in 
(\ref{P1P0G})  is actually $-15\zeta[3]/(128\pi^6)\,x^6 \simeq -1.47\,10^{-4}\, x^6$ (which is indeed exactly the last term 
of Eq.~(5.8) in \cite{phi4N} in the large-$N$ normalization of $b_0$). One should be evidently cautious not to use (\ref{P1P0G})
beyond its range of validity, typically for $m \gg T$, in such a case one rather solves 
the exact mass gap, Eq.~(\ref{opt0Tex})).\\
One may also easily derive the perturbative expansion of the pressure, 
which reads to few first orders
\bea
&\frac{P^{(1)}}{P_0} \simeq 1 -\frac{5}{4}\alpha +\frac{5}{3} \sqrt{6} \alpha^{3/2} +\frac{5}{4}
(L_T-6) \alpha^2 -\frac{5}{2}\sqrt{6} \left (L_T-\frac{3}{2}\right )\alpha^{5/2} -
\frac{5}{4} \left [L_T\left (L_T-12\right )+6\right ]\alpha^3 \nn \\
&+\frac{5}{32}\sqrt{6}\left [20 L_T\left (L_T-3\right )+9\right ]\alpha^{7/2}
+\frac{5}{4}L_T\left [L_T\left (L_T-18\right )+18\right ] \alpha^4+{\cal O}(\alpha^{9/2})\;\;,
\label{P1P0exp}
\eea
where $\alpha\equiv b_0 \lambda$.
We remark again that the one- and two-loop standard perturbative terms for 
the physical massless pressure~\cite{Trev,SPT3l,Pexpg7} would be reproduced by Eq.~(\ref{P1P0exp}) if effectively replacing, 
following ref.~\cite{2PI}, $b_0\to b_0/3=1/(16\pi^2)$. But it is easily seen that  
the correct coefficients of the leading logarithm terms of the pressure at  arbitrary order $n$: $\ \alpha^n\,ln^{n-1}(\mu)$ and $ \alpha^{n+1/2}\,\ln^{n-1}(\mu)$ (appearing first at 
three-loop order $\lambda^2$), are given by $(b_0 \lambda/3)(b_0 \ln (\mu)\, \lambda)^{n-1}$ 
with the correct $b_0$ (comparing with Ref.~\cite{Pexpg7} where the calculation was performed up to order 
$\alpha^4 \ln \alpha$ using RG techniques). Thus taking $b_0/3$ is not consistent beyond the first two perturbative terms.
Eq.~(\ref{P1P0exp}) being RG invariant by construction correctly reproduces the leading
logarithm structure to all orders beyond two-loop order.  
We stress also that Eq. (\ref{P1P0exp}) correctly
reproduces all perturbative terms of the large-$N$ result in \cite{phi4N}, when taking the correct value of the large-$N$ $b_0$. 
Since our expressions (\ref{P1P0}), (\ref{P1P0G}) are valid for arbitrary $N$ we can in principle follow continuously
the pressure from large $N$ to $N=1$, and while doing this there is no reason to abruptly modify the correct $b_0(N)$ 
to some other ``effective'' $b_0$ value. It is useful at this stage to compare this behavior with 
the SPT pressure up to two-loop or higher order~\cite{SPT3l}, basically 
build on taking $a=1/2$ in (\ref{subst1}): it does reproduce the
coefficients of the standard perturbative pressure up to second $\alpha^{3/2}$ order, but not the correct 
leading logarithm coefficients at order $\alpha^2$ and beyond, as a consequence of missing RG invariance (it would need
to rescale $\ln\mu \to 3\ln\mu$ to reproduce those logarithms). So, 
the scale dependence of the SPT pressure is unmatched at order $\alpha^2$ and beyond if using the standard running coupling with $b_0$.\\
Thus, keeping the correct $b_0$, which is compelling in our RG-based approach, the optimized results Eqs.~(\ref{P1P0G}) and (\ref{P1P0exp}) 
differ from the first two terms of the standard (massless) perturbative pressure for 
small coupling values by  $\lambda(\mu_0) \to \lambda_{pert}(\mu_0)/3$, 
for the canonically normalized coupling of the $N=1$ scalar model. 
But this is not a problem, simply a different calibration: 
the exactly scale-invariant RGOPT pressure (\ref{P1P0G}) only depends on the single coupling $G$ or equivalently $\lambda(\mu_0)$,  
still an arbitrary parameter at this stage, since the model is not fully specified by any data fixing a physical input scale, $\mu_0$.
So the pressure as a function of this coupling has a limited physical meaning. When going to higher loops, since the 
nonperturbative RGOPT approximations resum more higher orders,  it is not surprising 
that they differ from standard perturbation when expressed in terms of the original perturbative coupling $\lambda(\mu)$,
and one expects to obtain a better approximation for large coupling. The only mandatory feature 
of any such approximation is certainly the Stefan-Boltzmann limit $P \to P_0$ for $\lambda\to 0$, trivially fulfilled by 
(\ref{P1P0G}). Indeed, those features do not contradict truly physical results, as
this apparent discrepancy disappears if expressing the pressure in terms of the {\em physical} mass: 
to see it, we solve Eq.~(\ref{OPT}) now reciprocally, for $\t \lambda(m)$, and replacing it in (\ref{P1P0}). It 
gives simply: 
\be
P^{(1)}/P_0 = 1-15x^2/(8\pi^2)+15x^3/(8\pi^3)+{\cal O}(10^{-4}x^6).
\ee
 But here   
$x= m/T$ is arbitrary as we already used (\ref{opt0T}) to fix $\t \lambda(m)$. Now, taking for $m$ the physical screening 
mass~\cite{mdeb} $m_D$ in Eq.~(\ref{mD1}), as easily checked it
exactly reproduces the first two terms of the standard physical pressure~\cite{Trev}. Thus if we would plot our results in terms of
the screening mass, 
$P(m^2_D/T^2)$, we would have very good agreement
with standard results for sufficiently small screening mass $m^2_D$, and deviations for larger $m^2$, 
but the study of scale dependence which is our main concern, specially for large coupling values, would be much more difficult. 
In the sequel we keep the results (\ref{P1P0G}) in terms of the 
$\ms$ coupling $\lambda(\mu_0)$, which scale dependence is well-defined at a given perturbative order,
since our aim is mainly to compare the scale dependence with other
results in the literature also mostly expressed in terms of the running coupling. \\
The exact expression for $P^{(1)}/P_0$, Eq.~(\ref{P1P0}), is plotted
in Fig. \ref{Popt0T} where the RGOPT result is compared 
with the standard perturbative expansion at order $\lambda \sim g^2$. 
The pressure for the rescaled coupling $\lambda\to\lambda/3$ is also shown on the same Figure just for the sake of illustration, 
which accordingly compares better with the standard perturbative pressure for small coupling values.
The improvement of scale
(in)dependence of RGOPT is once again drastic at this one-loop order:
using the exact one-loop resummed coupling, Eq.~(\ref{lamrun}),  the RGOPT pressure is exactly 
scale-invariant, which is obvious in Eq.~(\ref{P1P0}) since $\t  m$ is itself
exactly scale-invariant and the combination $1/\lambda(\mu)+b_0 L_T$ too, as discussed above. 
This feature is well  illustrated by Fig. \ref{Popt0T} 
by comparing the RGOPT with the standard perturbative pressure at one-loop which has a 
notoriously large scale dependence. \\
To conclude this section we stress that all the previous RGOPT one-loop results, reproducing among other things the exact large-$N$ results,
only rely so far on the very simple massive one-loop
free energy graph and the knowledge of the first order RG coefficients $b_0$, $\gamma_0$. But these results are not too surprising 
since the RG properties, if fully exploited, involve informations on all orders ``daisy'' and ``super-daisy'' foam graphs like
those explicitly resummed in the large-$N$ limit in \cite{phi4N}.
\section{2-loop, ${\cal O}(\delta)$}
We now switch to the two-loop order,  
thus incorporating all the terms in Eq.~(\ref{F02l}), adding the subtraction
terms in Eqs. (\ref{sub}) and (\ref{s0s1}), and then  performing the $\delta$-expansion 
consistently to order $\delta$ before setting $\delta=1$. For simplicity we first consider the
minimal prescription taking only $s_1\ne 0$, and will also later  consider 
the relevant changes if the $s_2$ term, formally of order $\lambda$, is also included.
The main novelty at two-loop order is that now the RG relation, Eq.~(\ref{RGred}), gives in general 
a nontrivial additional constraint, that can be used alternatively to the OPT equation, or combined
with the latter to completely fix $m$ {\em and} $\lambda$ in terms of the only remaining free parameter, $\mu/T$ 
(apart from the overall dimensional dependence in $T$). 
This is another difference with standard OPT or SPT/HTLpt, in which 
the coupling remains undetermined and the generally adopted prescription  is 
to take its perturbative value as a function of $\mu/T$ and a reference coupling $\lambda(\mu_0)$ value. 
In our case, as explained previously one of the RG solutions is matching 
this standard perturbative behavior for $\lambda\to 0$,
but for moderate or larger coupling values it will give a nonperturbative dependence.  
One may thus follow at this stage different possible prescriptions: either, one can use
any of the two OPT and RG equations, to be solved customarily for $\t m(\lambda)$, next
using a two-loop order running coupling, in order to compare with other resummation methods. 
Alternatively, one may consider the full RG and OPT combined solutions.
Whatever way, neither the optimized mass $\t m$ solution of Eq.~(\ref{OPT}), nor the optimized
coupling $\t \lambda$ when combining the former with the constraint given by Eq.~(\ref{RGred}), 
have intrinsic universal physical meaning. 
Both should better be viewed as intermediate stage values, to be used only within the physical quantities such
as the pressure $P(\t m,\t \lambda,\mu/T)$.  In particular
there is no contradiction between the ``fixed" optimized coupling $\t \lambda$ and 
the standard running coupling, obtained from a different (standard) perturbative RG equation.\\

Like for the one-loop approximation,  we can express all two-loop RGOPT results 
in terms of the one-loop self-energy defined in Eq.~(\ref{Sigma}).
After some algebra, the ${\cal O}(\delta), \delta\to 1$  free energy takes
a compact form:
\be
{\cal F}_0 = -\frac{m^4}{(4\pi)^2} \left (\frac{1}{3b_0\lambda}+\frac{s_1}{3}+s_2\,\lambda \right ) +\frac{1}{2}
\int_{p,R} \ln (p^2+m^2) -\frac{m^2}{\lambda}\left (\frac{2\gamma_0}{b_0}\right )\:\Sigma_R +
\frac{1}{2\lambda} \Sigma_R^2 \;,
\label{del2compact}
\ee
where the index ``$R$'' in the integration means taking the finite part of this 
already renormalized expression. We also 
 kept as much as possible a general dependence on RG coefficients.
The first three terms originate from the subtraction terms $s_i$  in Eqs.~(\ref{sub})-(\ref{s0s1}).
Notice also the different coefficient $1/(3b_0)$ as a result of 
expanding to ${\cal O}(\delta^1)$, instead of previous $1/(2b_0)$ at one-loop $\delta^0$ order. As already mentioned we first consider 
for simplicity $s_2=0$ in the sequel, while the effects from $s_2\ne 0$ (that incorporates a RG part of the three-loop contributions) will be discussed later.\\
Next after straightforward manipulations the OPT Eq.~(\ref{OPT}), and the {\em reduced} RG operator, Eq.~(\ref{RGred}), acquires 
a compact neat form:
\be
f_{\rm OPT}(m,\lambda,\frac{\mu}{T})= \frac{2}{3} h \left (-s_1-\frac{1}{b_0\lambda} \right ) +\frac{2}{3} S +\Sigma^\prime_R 
\left (S-\frac{1}{3\lambda}\right )\equiv 0\;,
\label{OPT2l}
\ee
\be
f_{\rm RG}(m,\lambda,\frac{\mu}{T})=  h \left[ \frac{1}{6}+\left (\frac{b_1}{3b_0}-S\right )\lambda \right ] + 
\frac{1}{2} \beta^{(2)}(\lambda) S^2 \equiv 0\;,
\label{RG2l}
\ee
with $h\equiv (4\pi)^{-2}$, $\beta^{(2)}(\lambda)\equiv b_0\lambda^2 +b_1\lambda^3$ is the
standard $\beta$-function restricted to two-loops, and recalling also that $s_1=-1$. We have defined
for convenience the reduced (dimensionless)
self-energy $S(m,\mu,T)\equiv \Sigma_R/(m^2\lambda)$ thus independent of $\lambda$, 
which makes the coupling dependence very transparent in Eqs.~(\ref{OPT2l}) and (\ref{RG2l}).
For completeness and further use below we also have from Eq.(\ref{Sigma}):
\be
\Sigma^\prime_R\equiv \frac{\partial\,\Sigma_R}{\partial m^2}= \lambda\left (S+m^2 \frac{\partial\,S}{\partial m^2} \right ) =
\gamma_0\,\lambda \,\left(\ln \frac{m^2}{\mu^2} -J_2(m/T)\right)\;.
\label{Sprime}
\ee
Note that in principle the reduced RG Eq.(\ref{RG2l}) is only valid 
when combined with Eq.~(\ref{OPT2l}) since the latter removes the $\partial_m$ part
of the complete RG operator in Eq.~(\ref{RG}). Clearly,  
 in the above normalization,   the complete RG Eq.~(\ref{RG}) reads
\be
f_{\rm RG\,full} \equiv f_{\rm RG}+2\gamma_m(\lambda) f_{\rm OPT} \equiv 0\;,
\label{RG2lfull}
\ee
where the anomalous mass dimension $\gamma_m$ 
(truncated at the two-loop order) was defined
in Eq.~(\ref{gamma}). Therefore, to obtain the most general solution
$\t m(\lambda)$ consistent with arbitrary coupling  values, Eq.~(\ref{RG2lfull})
should be solved.
As we shall see below, the solutions $\t m_{\rm RG}(\lambda)$ and $\t m_{\rm RG\, full}(\lambda)$ 
are very close for sufficiently small $\lambda$ but can depart substantially from each other
for arbitrarily large coupling. The advantage of using the reduced RG operator is that  
the solution can be more easily found  when looking for the intersection between the two RG and OPT solutions $m_{\rm OPT}(\lambda)$
and $m_{\rm RG}(\lambda)$. \\
Before proceeding, another digressing remark is  that the standard OPT/SPT would
correspond to a much simpler OPT equation than Eq.~(\ref{OPT2l}), since  
in particular the first two (subtraction) terms $s_0, s_1$ would be absent, resulting in a
simple OPT self-consistent solution: $\tilde m^2 \equiv \Sigma_R$. Moreover the modified 
RG Eqs.~(\ref{RG2l}) or (\ref{RG2lfull}) are usually never considered
within the standard OPT/SPT or HTLpt applications. The fact that the 
OPT and RG relations, Eq.~(\ref{OPT2l}) and Eq.~(\ref{RG2l}), are more involved is expected 
since the one-loop RGOPT already gives nontrivial results qualitatively similar to 
two-loop standard OPT/SPT. Accordingly, the relative complexity of RGOPT equations
at two-loop is due to the more information they carry on higher RG orders, and is a price for a more efficient and RG-consistent 
resummation procedure.  At this stage the OPT and RG Eqs.~(\ref{OPT2l}), (\ref{RG2l}) could be solved exactly 
for $\lambda(m,\mu/T)$, being respectively 
quadratic and cubic algebraic equations in $\lambda$. But to compare with most results in the literature 
it is more customary to rather solve for a mass gap $m(\lambda,\mu/T)$, to obtain in a next stage the pressure or other 
thermodynamical quantities as a function of the coupling.\\
Considering first the (reduced) RG Eq.~(\ref{RG2l}), it is best solved in a first stage 
as a simple quadratic equation for $S(m,\mu,T)$ whose  mass gap solutions are:
\be
S^{\mp}_{\rm RG}(\lambda) \equiv \frac{\Sigma_R}{\lambda\, m^2}= \frac{1\mp \sqrt{1-(1+\frac{b_1}{b_0}\lambda)(1+\frac{2b_1}{b_0}\lambda)}}
{3\lambda (1+\frac{b_1}{b_0}\lambda)} \;\;,
\label{Srgsol}
\ee
 where again the explicit RG dependence has been kept for generality.
Note that the coupling dependence is entirely contained in the RHS of Eq.~(\ref{Srgsol}) since by definition
$S$ does not depend on $\lambda$, prior to using the RG Eq.~(\ref{RG2l}). 
 Just to see where we stand before considering the more involved exact mass gap solutions at two-loop order,
let us consider Eq.~(\ref{Srgsol}) by crudely neglecting the two-loop $\beta$-function coefficient,
$b_1=0$, or equivalently taking the leading term when expanding (\ref{Srgsol}) for $\lambda\to 0$ . 
It gives immediately $S_{\rm RG}(\lambda) =1/(3\,\lambda)$ which, recalling that $S(m,\mu,T)\equiv \Sigma_R/(\lambda\,m^2)$ and using (\ref{Sigma}),
is nothing but the one-loop mass gap Eq.~(\ref{solopt0}) consistently recovered. 
Once we use a nonzero $b_1\ne 0$ in Eq.~(\ref{Srgsol}), at it should at two-loop RG order, one could consider a perturbative expansion of 
Eq.~(\ref{Srgsol}) to gradually include higher perturbative order corrections to the one-loop solution (\ref{solopt0}), but it is algebraically not more complicated 
to solve Eq.~(\ref{Srgsol}) exactly.
Now in fact Eq.~(\ref{Srgsol}) also reflects a possible  
complication appearing at two-loop order for large coupling values, due to the two-loop ultraviolet fixed point (UVFP) at
$\lambda =-b_0/b_1$ since $b_1 <0$, see Eq.~(\ref{rgcoeff}). This purely perturbative UVFP is totally
spurious, not only since it disappears at three-loop level (where the next coefficient $b_2$ is 
positive\cite{RGphi4loop} and large enough so that possible non-trivial fixed points are complex) but more generally 
since the existence of nonperturbative UVFP is excluded by the numerical evidence 
for the triviality~\cite{triviality} of the $\phi^4$ theory. Nevertheless, since the
RGOPT construction basically relies on perturbative RG properties, one may worry 
that some of our two-loop results could  
be affected, if driven by this spurious UVFP. Indeed the $(+)$ solution in Eq.~(\ref{Srgsol}) is singular at 
the fixed point value of $\lambda$, which means that $\t m^2/\Sigma_R(\t m^2)\to 0$ i.e. $\t m\to 0$, 
while the $(-)$ solution is regular: $S^-_{\rm RG}(-b_0/b_1) \sim b_1/(6b_0)$, which means that it is a priori the solution not wrongly driven by the
UVFP.
Thus, in the sequel we should be careful to identify any behavior that could be an artifact of this perturbative
fixed point. 
Now,  in terms of the rescaled coupling $\lambda=24g^2$,
the UVFP is at $g\sim 1.866$, and the maximum of the $\beta$-function (beyond which the coupling is driven
to a really wrong behavior), 
 is at $\lambda=-2b_0/(3b_1)$
i.e. $g\simeq 1.524$. Both values are to be considered very large
couplings, where the validity of a resummation
procedure is anyhow questionable.
Therefore, as long as one stays safely below these large coupling values,
say not too much above $g\simeq 1$ in practice, 
 our results should remain valid. Moreover, here we are basically focusing 
on the RG/scale invariance issues in very general terms, rather than on the 
peculiar nonperturbative dynamics of the $\phi^4$ model,  which is beyond the present scope. 
Another property to notice is that the exact RG solutions, Eq.~(\ref{Srgsol}), become
complex at a coupling $\lambda_c= -3b_0/(2b_1) = (4\pi)^2 27/34$ corresponding to  
$g\simeq 2.28$, thus irrelevant since located beyond the fixed point anyway. 
Yet one should keep in mind that independently of the presence of non-trivial perturbative
fixed points, complex optimized RG solutions are unavoidably   
expected to occur at some higher perturbative order
from exactly solving the OPT and RG equations, as discussed above. 
\subsection{$T=0$}
%
\begin{figure}[!]
\epsfig{figure=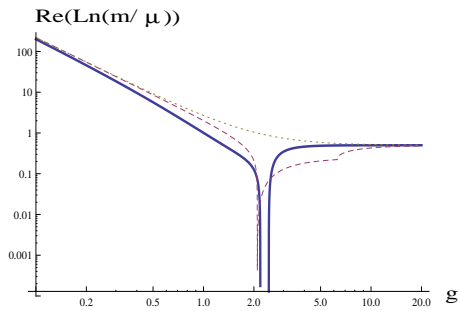,width=6cm,height=4cm}
\caption[long]{$T=0$ two-loop ($\delta^1$) order OPT (dashed) and RG (thick) (real parts of)
solutions $\ln \frac{m}{\mu}(g)$, $g\equiv \sqrt{\lambda/24})$. Also shown is the one-loop solution (dotted)
from Eq.~(\ref{opt0}).}
\label{LoptLrg}
\end{figure}
To get some more feeling we first explore the $T=0$ case, which is much simpler
since both RG and OPT equations can be solved analytically, {\it e.g.} for $\ln m/\mu$ in terms
of the (rescaled) coupling $g\equiv \sqrt{\lambda/24}$ (or its reciprocal function $\lambda(\ln m/\mu)$). 
The (real parts) of the OPT and RG exact solutions, expressed  as $\ln [m(g)/\mu]$, are plotted
in Fig. \ref{LoptLrg}. We also show 
for comparison the one-loop solution (dotted) from Eq.~(\ref{opt0}).
The physical branch solutions are clearly identified, i.e. those matching standard perturbation
for $\mu\ll m$ with  fixed $m$:
$\ln m/\mu\to +\infty$ for $\lambda\to 0^+$ (see also the discussion after Eq.~(\ref{solopt0})).
One also sees the asymptotic value $\ln m/\mu \simeq 1/2$ reached for large coupling $g$ consistently
with Eq.~(\ref{opt0}).
The two-loop OPT (dashed) physical branch becomes complex for $g \gsim 2.094$ (where there is
a corresponding bifurcation on the figure), in fact very close above the 
intersection between the two (real) physical branch solutions, occurring at $\t g \simeq 2.08, 
\ln (\t m/\mu) \simeq 0.083$, thus at $\t m/\mu$ close to 1 and 
a quite strong coupling value. The OPT branch is real again at about $g\gsim 6$ (where the dashed curve
shows a little bump). The RG physical branch becomes also complex at a slightly
higher $g\simeq 2.28$ value, as already noted above after Eq.(\ref{Srgsol}).
There are also two other combined RG and OPT solutions
(intersections) sitting on the complex branches, see Fig. \ref{LoptLrg}, but these
are to be considered unphysical solutions since not 
connected with the perturbative 
branches. One can also note that the RG branch has a pole behavior around $g \simeq 2$,  which
is a consequence of the above discussed perturbative UVFP at 
$g=\sqrt{-b_0/(24b_1)}\simeq 1.866$.\footnote{The pole of $\ln m/\mu$ is not exactly
at the fixed point of $\beta(\lambda)$ due to the mass-dependence entering the RG Eq.~(\ref{RGred}).}
The reciprocal function $g(\ln m/\mu)$ would show a frozen behavior at $g\simeq 2$. In fact, 
we stress that the behavior around $g\simeq 2$, including the solutions becoming complex, is all 
driven by this naive perturbative two-loop UVFP, 
so that one should simply not trust what happens for $g$ close to those values, say for $g\gsim 1$
to be on the safe side. (Notice however that on the figure the
one-loop-like behavior is recovered for much larger $g$ values). In particular, the above mentioned real OPT and RG 
intersection solution at $\t g\simeq 2.08$ is beyond the UVFP, thus very untrustable. From examples in other theories~\cite{rgopt_alphas}, we
expect that at higher orders the RGOPT intersection solution may decrease below the UVFP and stabilize to a more reasonably
perturbative value.\\
Switching on the thermal contributions 
modifies coefficients of the relevant RG and OPT equations, which will 
result in real intersecting points for the RG and OPT solutions 
with somewhat lower coupling values $g\sim 1$ for generic $\mu/T$ values 
as we examine in next subsection below\footnote{When complex solutions occur on physical branches, 
one may recover real solutions by performing a perturbative scheme change, 
as done in \cite{rgopt_alphas}. But  
this more involved course of action can be avoided in the $\phi^4$ case, at least at the two-loop level. We anticipate
however that for thermal QCD, RGOPT will unavoidably gives complex solutions, 
mainly due to the opposite signs of the $b_0, b_1$ coefficients 
due to asymptotic freedom.}. 
\subsection{$T\ne 0$}
\begin{figure}[h!]
\epsfig{figure=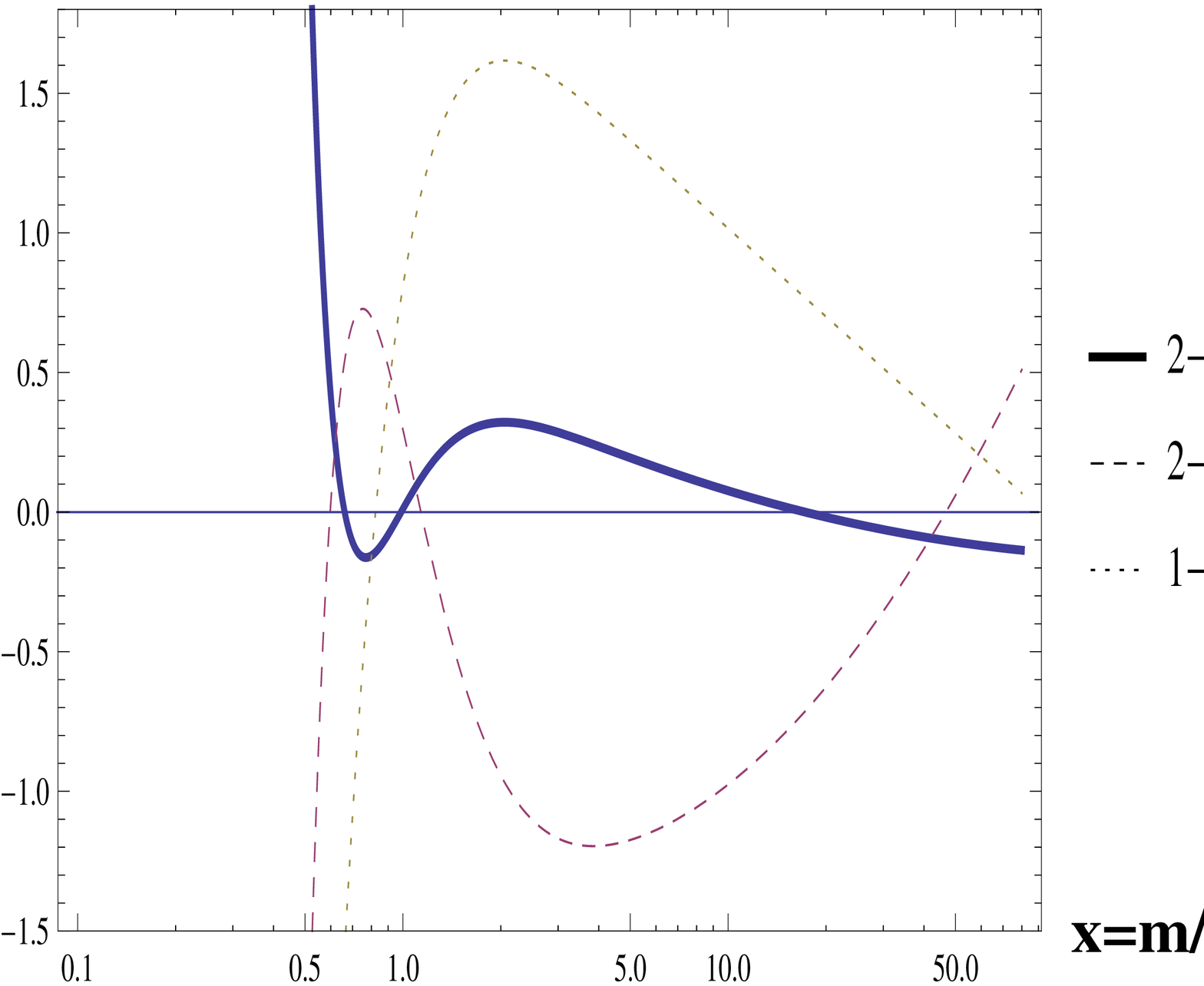,width=8cm,height=5cm}
\caption[long]{Roots of the two-loop ($\delta^1$) exact OPT Eq.~(\ref{OPT2l}) and RG Eq.~(\ref{RG2l}), 
as compared with one-loop OPT Eq.~(\ref{opt0Tex}) for  $g=\sqrt{\lambda(\mu_0)/24}=1$, $\mu_0=2\pi T$. The $y$-axis values
are those of $f_{\rm OPT}$ Eq.~(\ref{OPT2l}), $f_{\rm RG}$ Eq.~(\ref{RG2l}), and  Eq.~(\ref{opt0Tex}) in convenient common units.}
\label{OPTRGex}
\end{figure}

Considering  now the thermal contributions,  one may
solve numerically Eq.~(\ref{Srgsol}) (or the full RG Eq.~(\ref{RG2lfull}))
and the OPT Eq.(\ref{OPT2l}), 
using the exact expression $S(m,T,\mu)$ from Eq.~(\ref{Sigma}), to obtain 
$x\equiv m/T$  as function of $\lambda(\mu)$ at some chosen scale $\mu$. 
Concretely,  to solve the OPT gap-equation exactly for arbitrary temperature, it is convenient to first solve Eq.~(\ref{OPT2l}) 
as a linear equation for $S(m/T,\mu/T)$ in terms of $\lambda$ and $S'(m/T,\mu/T)$, giving trivially
\be
S\equiv\frac{\Sigma_R}{\lambda\,m^2} = \frac{1}{3\lambda}-\frac{1}{8\pi^2\,(2+3\lambda\,S')}\;,
\label{Soptex}
\ee
to be then solved numerically as a mass gap $\t m(\lambda,\mu/T)$ as function 
of the coupling and scale, using the expressions of $S, S'$ in Eqs~(\ref{Sigma}), (\ref{Sprime}). 
In the right hand side of Eq.~(\ref{Soptex}), taking only the first term, 
dominant for $\lambda\to 0$, one immediately recovers again the one-loop RGOPT mass gap solution Eq.~(\ref{solopt0}) (just like 
for the above discussed RG mass gap (\ref{Srgsol}) when taking $b_1=0$), while the second term of Eq.~(\ref{Soptex}) clearly gives higher order
corrections if seen as a perturbative expansion. (However, for large coupling values of order $g=\sqrt{\lambda/24}\sim 1$ such a perturbative expansion of Eq. (\ref{Soptex})
would not give a very accurate mass gap solution so it is better to solve it exactly numerically).\\ 
The exact OPT Eq.~(\ref{Soptex}) and RG Eq.~(\ref{Srgsol}) are illustrated with their roots for a rather strong coupling value $g=1$ in 
Fig. \ref{OPTRGex}.
As one can see, in general both the two-loop order OPT and RG equations have three real solutions, until
two solutions become complex (conjugates), which happens for $g\gsim 2.09$ and $g \gsim 2.28$ respectively for the OPT and RG equations. 
Since this is well beyond the fake perturbative UVFP, we cannot trust the detailed consequences near such large coupling values.
For more moderate coupling values as illustrated in Fig. \ref{OPTRGex}, there is one RG and one OPT solution 
with very large $m/T\gg 1$, in fact behaving for small $\lambda$ as $\t m \sim e^{1/(b_0\lambda)}$ as one can easily trace even from the one-loop
mass gap Eq.~(\ref{opt0Tex}): for $m\gg T$ the $T^2 J_1(m/T)$ term in (\ref{opt0Tex}) becomes negligible and one simply recovers
the $T=0$ solution for $m(\lambda)$. But for $T\ne 0$ this solution does not have the property of a thermal mass, $m\to 0$ for $T\to 0$,   
so that the other roots are the physically relevant ones. 
The behavior of the two intermediate and large values roots is 
qualitatively similar to the one-loop OPT also illustrated and large-$N$ mass gap solution~\cite{phi4N}, indeed recovered as above explained at one-loop order. 
What is new as compared to one-loop order and seen on Fig. \ref{OPTRGex} 
are the two extra roots with the lowest $x=m/T$ values both for the OPT and RG equations. 
Concerning the RG root with the lowest $x\simeq 0.7$ value, it is easily traced to be the one driven by the UV fixed point: 
for $\lambda\to -b_0/b_1$,  it gives $x\to 0$, and we should reject it accordingly. 
The other OPT root with lowest $x\simeq 0.6$ value is more special: it is more easily analyzed by solving the exact OPT Eq.~(\ref{OPT2l})
for $\lambda(m/T)$, which is a simple quadratic equation. It can also be seen to correspond to 
the perturbatively odd situation where the second term in Eq.~(\ref{Soptex}) dominates over the first term. 
Then matching with the perturbative behavior, 
it gives an ``ultrasoft'' mass $m/T \simeq (\pi b_0/2)\lambda +{\cal O}(\lambda^2)$ for $\lambda\to 0$, which contradicts 
the expected behavior of a thermal mass $m\sim \sqrt{\lambda}\,T$ on general grounds, that was indeed found at one-loop order. At higher orders
this optimized mass has a standard power series in $\lambda$.  
So we consider this solution as a spurious unphysical one, an artifact of the more involved two-loop OPT equation.
Therefore at two-loop order we identify unique physical OPT and RG solution, 
which are real and positive for all relevant coupling values and compatible with the
perturbative behavior of a thermal mass, corresponding to the two intermediate value roots near $x\simeq 1$ illustrated on Fig. \ref{OPTRGex}. 
Once correctly identified, those physical solutions
have a qualitative behavior not drastically different from the one-loop order OPT solution of (\ref{opt0Tex}), 
apart from the more involved algebra. 
Note also that solving Eqs.~(\ref{Srgsol}) or (\ref{Soptex}) for
$\t \lambda(m,\ln(\mu/T))$ one can check analytically that for fixed $m/T$ values, the physical 
$\lambda(\mu/T)$ solution decreases logarithmically in the two extreme limits $\mu\ll 2\pi T$ and $\mu \gg 2\pi T$, having a maximum
in between, a behavior qualitatively quite similar to the large-$N$ or one-loop case.\\
Alternatively for any solutions, as long as $x\lsim 1$, 
one can solve analytically both RG and OPT equations 
but using the high-$T$ approximation (\ref{J0exp}) and derivatives for the relevant thermal 
integrals, giving 
\be
S(x\equiv m/T\lsim 1) \simeq -\frac{1}{(4\pi)^2}\left (L_T+\frac{2\pi}{x}-\frac{2\pi^2}{x^2}\right )\;\;.
\label{ShighT}
\ee
In this way one obtains respectively quartic RG and cubic 
OPT equations in $x=m/T$, which are not particularly telling but gives algebraic solutions. 
We checked that all the approximate high-$T$ RG solutions for $\t m$ and the pressure $P/P_0$
for relevant scale values $\pi T<\mu <4\pi T$
are excellent, departing below $0.1\%$ from the exact $T$-dependent solutions, 
at least up to a value of the (rescaled) coupling $g\equiv \sqrt{\lambda/24} \sim 1.5$,
simply because $\t x$ remains always lower than about $\sim 1$. Concerning the OPT solutions,
they can give $\t x >1$ at large coupling $g \gsim 1$, particularly for the higher scale 
choice $\mu=4\pi T$, in which case the high-$T$ approximation starts to fail and 
we better use the exact $T$-dependent numerical solutions~\footnote{Such remarks are important 
to keep in mind in view of possible applications to thermal QCD, 
for which beyond the one-loop level only high-$T$ expansion results are 
available~\cite{HTLPT3loop}.} of Eq.~(\ref{Soptex}). 
\subsubsection{Comparison with standard perturbation theory}
\begin{figure}[h!]
\epsfig{figure=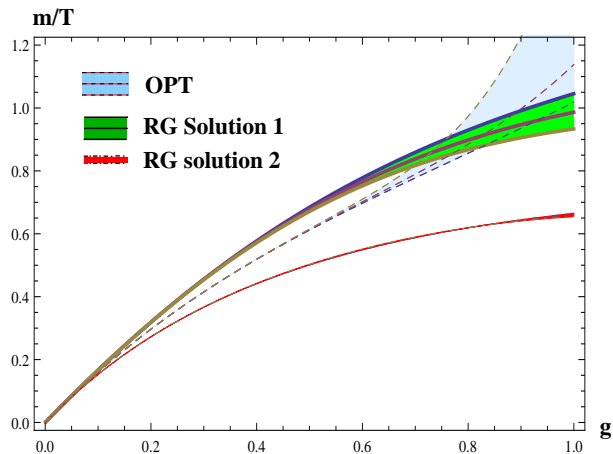,width=8cm,height=6cm}
\caption[long]{Two-loop ($\delta^1$) optimized mass solutions obtained from
OPT Eq.(\ref{OPT}) (dashed, light blue on line) and RG 
Eq.(\ref{RGred}) two solutions: thick lines (green on line) and dot-dashed (red on-line) respectively, as function of $g=\sqrt{\lambda(\mu)/24}$ 
with scale dependence $\pi T < \mu < 4\pi T$.}
\label{Mrgopt2Tall}
\end{figure}

\begin{figure}[!]
\epsfig{figure=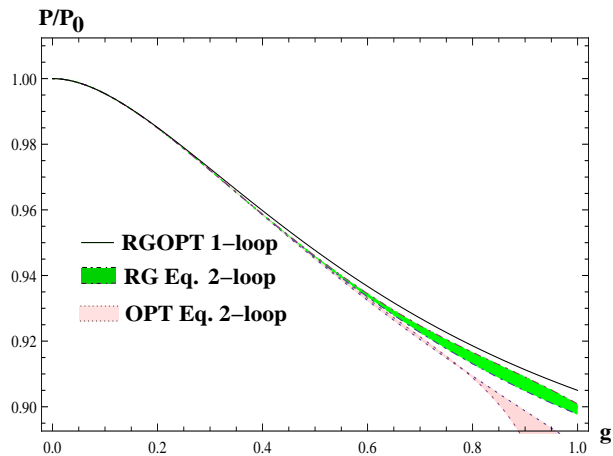,width=8cm,height=6cm}
\caption[long]{Two-loop pressure $P/P_0(g\equiv \sqrt{\lambda/24})$ obtained from the full RG solution of (\ref{RG2lfull}) (dot-dashed, green) and 
obtained from the OPT solution (\ref{OPT}) (dotted, light red), with scale dependence between 
$\mu=\pi T$ and $\mu=4\pi T$. The RGOPT one-loop pressure (black line) is also shown for comparison. }
\label{Prgvsopt}
\end{figure}

The OPT physical solution has the following perturbative expansion
\be
\frac{{ \t m}^{(2)}_{\rm OPT}}{T} \sim \pi \sqrt{\frac{2}{3}}\sqrt{b_0 \lambda} -\pi b_0 \lambda +\frac{3}{128\pi^2\sqrt 2}(5-2L_T) \lambda^{3/2}
-\frac{9}{1024\pi^3}(3-4L_T) \lambda^2 +\cdots 
\label{solopt1Tpert}
\ee
consistently with the first two terms of Eq.~(\ref{solopt0Tpert}). 
 Concerning the terms of order $\lambda^{3/2}$ and $\lambda^2$, the (leading) logarithms coefficient of $L_T$ are the same with respect to the
one-loop expansion  Eq.~(\ref{solopt0Tpert}), as it should be  to all orders. The differences appear only in the constant terms, 
due both to $s_1\ne 0$ and other terms of order $\lambda^0$ in the original expression of the free energy (\ref{del2compact}). 
Note that the relative weight of $s_1$
is changed at order $\delta^1$, with $s_1/3$ in (\ref{del2compact}). Quite similarly, incorporating $s_2\ne 0$ gives
differences only visible at order $\lambda^3$ not given in Eq.~(\ref{solopt1Tpert}), 
so it has very little effect at least on the perturbative mass expansion.
Plugging this in the pressure expression and
expanding perturbatively (while
keeping the coupling free for the moment) we obtain  the ratio of the two-loop RGOPT 
pressure to the ideal gas one $P_0=\pi^2 T^4/90$ as:
\be
\frac{P^{(2)}_{\rm OPT}}{P_0} \sim 1 -\frac{5}{4}\alpha +\frac{5}{3} \sqrt{6} \alpha^{3/2} 
+\frac{5}{4}(L_T-20/3) \alpha^2 -\frac{5}{2}\sqrt{6} (L_T-13/6) \alpha^{5/2} 
-\frac{5}{4} [L_T(L_T-40/3) +44/3]\alpha^3
+{\cal O}(\lambda^{7/2}) \;, 
\label{PP02lexp}
\ee
where $\alpha\equiv b_0 \lambda$, giving the same first two perturbative terms as the RGOPT one-loop result Eq.~(\ref{P1P0exp}). 
We remark also, just like in the one-loop case, that if rather solving either the OPT or the RG Eqs. (\ref{OPT2l}), (\ref{RG2lfull}) reciprocally
for $\lambda(m)$, and then taking for $m$ the physical screening mass Eq.~(\ref{mD1}), one consistently recovers the standard perturbative expression
of the physical pressure up to two-loop order. 
Remark  that the effect of incorporating $s_2\ne 0$ instead of the simplest choice with only $s_1$, 
appears first in the very last non-logarithmic term of order $\t m^4\lambda \sim \lambda^3$ in (\ref{PP02lexp}), 
replacing $44/3\to 637/45 -4\zeta[3]/5$,
which has very little effects upon the final numerical results as long as $\lambda$ remains reasonably perturbative. 
This is completely expected since perturbatively $m^4 s_2 \lambda$ in (\ref{sub}) 
is of lowest perturbative order $\sim \lambda^2 s_2\lambda\sim s_2 \lambda^3$. \\
Alternatively, we also consider the solution $\t m(\lambda,\mu/T)$
obtained from the (full) RG Eqs.~(\ref{RG2l}) and (\ref{RG2lfull}). 
The latter equations can be solved
for the exact $T$-dependence first as quadratic equation for $S(m/T,\mu/T)$ 
and then numerically for $m(\lambda,\mu/T)$. But as mentioned above the high-$T$
approximation is excellent in this case for all relevant coupling values. The 
 high-$T$ approximated RG equations gives two negative and two positive $\t m /T$ solutions,
the latter having both the correct perturbative behavior, but with one 
saturating faster for large coupling, which is illustrated as ``solution 2'' in Fig. \ref{Mrgopt2Tall}. 
In Fig. \ref{Mrgopt2Tall} we also show the scale
dependence in the range $\pi T < \mu < 4\pi T$ for all solutions $\t x\equiv \t m(g,\mu)/T$. 
The ``lower" values RG solution 2 happens to reach a maximum and then decreases 
towards zero for large coupling $g\sim 1.86$ (not shown in the figure), 
an odd behavior which is in fact completely driven by the perturbative two-loop UVFP. 
Incidentally this solution exhibits an extremely small, almost totally negligible scale dependence
up to $g\sim 1$, as can be seen in Fig. \ref{Mrgopt2Tall}, in agreement with what is intuitively expected
near a RG fixed point. The corresponding pressure, if plugging this solution
within (\ref{del2compact}), has even smaller scale dependence. Although we shall simply
discard this solution, since the naive two-loop UVFP contradicts 
the genuine nonperturbative dynamics of the $\phi^4$ model, it is worth remarking 
that the corresponding RGOPT solution faithfully reflects the quasi scale invariant behavior of the UVFP. 
The remaining (then unique physical) RG solution 1 exhibits a more pronounced scale dependence as seen in 
Fig. \ref{Mrgopt2Tall}, which will be further explained below. It
has a perturbative expansion for small coupling with the first two order terms identical with 
the other OPT solution above, Eq.(\ref{solopt1Tpert}):
\be
\frac{{ \t m}^{(2)}_{\rm RG}}{T} \sim \pi \sqrt{\frac{2}{3}}\sqrt{b_0 \lambda} -\pi b_0  \lambda 
+ \frac{(13-12 L_T)}{256\pi^2\sqrt 2}\lambda^{3/2} +\frac{9}{4096\pi^3}(5+16L_T) \lambda^2
+\cdots 
\label{solRG1Tpert}
\ee
where one can easily check that the coefficients of $L_T$, thus of $\ln \mu$, are identical, 
which should be the case to all orders consistently with RG invariance properties. \\
The perturbative expansion of $P/P_0$ is identical to (\ref{PP02lexp}) for the first few orders 
and for $L_T$ coefficients at all orders if we use instead the $\t m(\lambda)/T$ solution, Eq.~(\ref{solRG1Tpert}). \\
Remark in Fig. \ref{Mrgopt2Tall} the intersections, for different $\mu$ scale values,  
between the RG and OPT solutions, which only exist for the higher 
of the two RG solution branches (namely the physical solution not influenced by the perturbative UV fixed point). 
Those intersections can be considered as the full RGOPT solution, which is unique for a given $\mu$ input scale. 
We shall come back later on those full RGOPT solutions, while for the sake of comparison with standard perturbation 
results we consider at the moment the RG and OPT solutions separately, as given functions of the coupling. \\
\begin{figure}[!]
\epsfig{figure=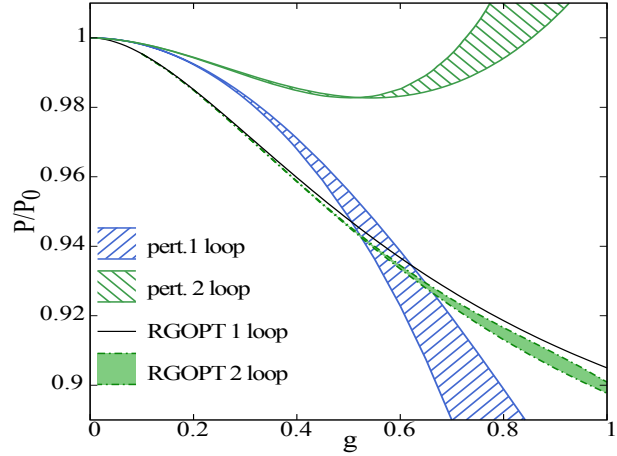,width=8cm,height=6cm}
\caption[long]{RGOPT one-loop (black line) and two-loop (dot-dashed, green on line) versus standard perturbative pressure $P/P_0(g\equiv \sqrt{\lambda/24})$ at 
one-loop and two-loop. The different bands give the scale dependence between 
$\mu=\pi T$ and $\mu=4\pi T$. NB: the RGOPT one-loop curve has actually zero thickness 
since it is exactly scale-invariant. }
\label{Prgopt1and2loop}
\end{figure}
In Fig. \ref{Prgopt1and2loop} we plot the exact two-loop RGOPT pressure $P/P_0(g\equiv \sqrt{\lambda/24})$ (dashed lines),
as obtained from the mass solution of the full RG Eq.~(\ref{RG2lfull}), 
compared with one-loop RGOPT and standard perturbative one- and two-loop results, 
with scale dependence in the range $ \pi T < \mu < 4\pi T$.
To study the scale dependence we use a standard running
coupling exact at two-loop orders. The generic exact expression for the two-loop running $\lambda(\mu)$, 
generalizing the exact one-loop running in 
Eq.~(\ref{lamrun}), can be expressed (see {\it e.g.} Ref. \cite{qcd2}) 
in terms of the (implicit) Lambert ``function''~\cite{Lambert} $W(x)\equiv \ln (x/W)$:
\be
\lambda(\mu) = \frac{\lambda(\mu_0)}{ f_W\left (\lambda(\mu_0),\ln \frac{\mu}{\mu_0}\right)}
\label{runex2l}\;\;,
\ee
with 
\be
f_W(\lambda,L) = 1-b_0 L \lambda + \frac{b_1}{b_0} \lambda \ln \left(f_W \:
\frac{1+\frac{b_1}{b_0} \lambda f^{-1}_W}{1+\frac{b_1}{b_0} \lambda}\right)
=-\frac{b_1}{b_0} \lambda \left\{1 +W\left[-\left (1+\frac{b_0}{b_1 \lambda}\right )\:
e^{-[1+\frac{b_0}{b_1 \lambda} (1-b_0 \lambda L) ]}\right] \right\}\;\;,
\ee
where $L\equiv \ln\mu/\mu_0$ and where we consider as usual the reference scale $\mu_0=2\pi T$.
Actually Eq.~(\ref{runex2l}) gives no visible difference (at least up to the relevant moderately large
coupling values here studied and the moderate 
range of scale variation) with a more handy 
perturbatively truncated  expansion at order $\lambda^3$:
\be
\lambda^{-1}(\mu) \simeq \lambda^{-1}(\mu_0) -b_0 L  -(b_1 L) \lambda -\left (\frac{1}{2} b_0 b_1 L^2\right ) 
\lambda^2 -\left (\frac{1}{2} b_1^2 L^2 +\frac{1}{3} b_0^2 b_1 L^3\right ) \lambda^3 +{\cal O}(\lambda^4)\;.
\label{run2lapprox}
\ee
The RGOPT improvement on scale dependence with respect to standard perturbative results
is drastic for the pressure,  as one can clearly see in figure \ref{Prgopt1and2loop}, 
although scale invariance is not exact at two-loop like it is at one-loop order: 
there is a moderate residual scale dependence, clearly visible 
in Fig.~\ref{Prgopt1and2loop} for moderately large (rescaled) coupling values $g \gsim 0.6$.
What is also clearly seen in Fig. \ref{Prgopt1and2loop} is the much better 
stability of the RGOPT results, since 
up to $g\simeq .5$, both the scale dependence and the difference between one- and two-loop RGOPT pressure
are hardly visible at the figure scale, in contrast with the already poorly convergent standard perturbative pressure for those values. 
 Accordingly, there are important extra cancellations of the scale-dependence happening when 
$\t m(\lambda,\mu)$ optimized solutions from Fig. \ref{Mrgopt2Tall} are plugged into 
the pressure expression $P/P_0(\t m,\lambda,\mu)$ in Eq.~(\ref{del2compact}). 
But we emphasize again that the optimized
masses $\t m^{(k)}_{\rm OPT}(\lambda,\mu)$ or $\t m^{(k)}_{\rm RG}(\lambda,\mu)$ at order-$\delta^k$ 
are intermediate, {\em unphysical} quantities, 
therefore not expected to be themselves scale-invariant in general (although at one-loop order, $\t m^{0}_{\rm OPT}(\lambda,\mu)$ 
in Eq.~(\ref{solopt0T}) is  exactly (one-loop) scale-invariant as explained above). 
Within the  
RGOPT procedure, RG-invariance is by construction required only for the pressure which represents the actual physical observable, 
resulting in Eqs.~(\ref{RG2l}) and (\ref{OPT2l}) optimizing the pressure. 
Accordingly, the further cancellation of scale-dependence of the two-loop OPT or RG masses 
$\t m_{\rm OPT}(\lambda,\mu)$ or $\t m_{\rm RG}(\lambda,\mu)$, 
once plugged into the pressure $P/P_0(m,\lambda,\mu)$, is expected from the resummation properties of the RGOPT procedure.  \\
The residual scale dependence of the two-loop RGOPT pressure 
is unavoidable due to the RGOPT construction being   
not exact but resulting from the optimization of actually two-loop restricted 
basic free energy by construction, where terms of order $\lambda^2$ and higher are truncated. (In contrast the one-loop results above were exactly
scale invariant because of the perfect matching of the exact one-loop running coupling with the mass gap (\ref{solopt0T})).\\ 
One can make those statements more precise by studying 
exactly at which perturbative order the scale dependence reappears: examining 
the perturbative expansion of the pressure above in Eq.(\ref{PP02lexp}) 
in which the coupling is replaced by its running
expression at truncated two-loop order (so only with $b_0, b_1$ dependence) 
Eqs.~(\ref{runex2l}) and (\ref{run2lapprox}), we have checked explicitly that the leading 
scale dependence reappears first at order $\lambda^3$:
\be
\frac{P_{\rm OPT}^{(2)}}{P_0} (\mu)|_{\rm leading} \simeq \frac{\lambda^3(\mu_0)}{16384\,\pi^6}\: 
\left (85 \ln \frac{\mu}{\mu_0}-2190.5 \right ) +{\cal O}(\lambda^{7/2})\;,
\label{2lmuremn}
\ee
{\it i.e.} formally at four-loop order, that is one order higher than the naively expected three-loop $\lambda^2$ from standard RG 
invariance properties\footnote{When rescaling the coupling as $\lambda\to 24g^2$, 
the leading remnant scale-dependence in Eq.(\ref{2lmuremn}) gives: 
$\sim (0.075 \ln \mu/\mu_0 -1.92) g^6$, where the lowest order coefficients  
are roughly of order ${\cal O}(1)$ in this normalization.}. 
 This feature can be anticipated in fact with a little insight without involved explicit high order expansion calculations: 
recall that the RGOPT construction including the subtraction terms in (\ref{sub}),
together with the RG invariance preserving interpolation (\ref{subst1}), guarantees that the two-loop free energy (\ref{F02l}) is RG invariant up to 
neglected three-loop terms, of order  ${\cal O}(m^4\lambda^2)$, but for {\em arbitrary} mass $m$. This implies that the mass gap
obtained either from the RG Eq.~(\ref{Srgsol}) or OPT Eq.~(\ref{Soptex}) has a remnant scale dependence appearing perturbatively at order 
$\t m^2/T^2 \sim \lambda (1+\cdots+\lambda^2\ln\mu)$, as could easily be checked by explicit expansion too. So it means that 
the lowest possible order at which a remnant scale dependence appears in the free energy (\ref{F02l}) is given by the first subtraction terms 
$-s_0\, m^4/\lambda$, giving remnant scale dependence at perturbative order $\lambda^3\ln\mu)$.
In contrast the remnant scale-dependence of the standard perturbative two-loop pressure appears
 at the expected order $\lambda^2$.\\
Using alternatively the RG solution $\t  m^{(2)}_{\rm RG}(\lambda)$, we find 
a residual perturbative scale-dependence reappearing also consistently at order $\lambda^3$ 
with the same coefficient of $\ln \mu$ than in Eq.(\ref{2lmuremn}), and a very similar
constant term, with $\sim -2190.5$ in Eq.~(\ref{2lmuremn}) replaced by $\sim -2009.9$. However, 
we stress that Eq.~(\ref{2lmuremn}) allows to understand the perturbative behavior of the remnant scale dependence, 
but does not properly reflect the actual {\em nonperturbative} scale-dependence of the full RG
or OPT solutions, which we used for the plots in Fig. \ref{Prgopt1and2loop}.  Indeed
the actual values and scale dependence of the pressure for relatively large coupling $g>0.6$ are very different 
than what would be obtained by a finite order perturbative truncation at order $\lambda^3$.
Accordingly, the moderate residual scale dependence seen on the plots for large $g$ values appears much better than what (\ref{2lmuremn})
would give, as a result of further nonperturbative cancellations among successive higher orders (of course higher orders of the
RGOPT resummation anyway based on a two-loop truncated free energy). Clearly when $g$ becomes of order $1$ $\t m/T$ is also of order 1 
so that the previous perturbative reasoning with (\ref{2lmuremn}) no longer apply and 
we simply recover that the scale invariance is guaranteed by construction
up to remnant terms of order $m^4 \lambda^2\ln \mu$.\\
As another illustration,  in Fig \ref{Prgvsopt} the exact pressure obtained from 
the OPT mass solution of Eq.~(\ref{OPT2l}) is compared with the one obtained from the RG equation.
Although both RG and OPT solutions 
have perturbative scale-dependence reapparing at the same $\lambda^3$ order, like in (\ref{2lmuremn}), 
the nonperturbative scale dependence of the OPT solution is almost negligible until $g\simeq 0.8$, while 
for larger coupling values it becomes more important than the one of the RG solution (as one can see also for the corresponding OPT mass
in Fig. \ref{Mrgopt2Tall}). 
We also remark in Fig \ref{Prgvsopt} that the pressure obtained from the RG mass remains closer to the one-loop RGOPT pressure 
for large coupling values, so that the convergence appears better. This, together with the better scale independence for large
coupling, is not very surprising since the RG mass solution originates
from the RG Eq.~(\ref{RG2lfull}) at the two-loop level. In contrast, the
OPT solution results solely from the mass optimization, Eq.~(\ref{OPT}), which incorporates RG properties more indirectly: it also exhibits good scale invariance 
 up to relatively large values $g\gsim 0.8$, beyond which it degrades quite rapidly. 
Also, the fact that the RG and OPT solutions are very close
to each other until relatively large coupling values $g\simeq 0.6$ shows an overall   
consistency, by quantifying the relatively small 
lack of exact RG invariance, since for an exact nonperturbative result the OPT and RG solutions would be identical. \\
\begin{figure}[!]
\epsfig{figure=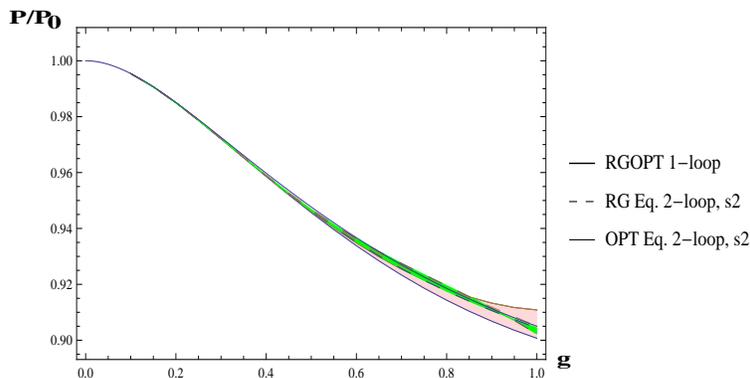,width=10cm,height=5cm}
\caption[long]{Same  as in Fig. \ref{Prgvsopt}, but with $s_2\ne 0$ in Eq.~(\ref{sub}). }
\label{P2loop_s2}
\end{figure}
We now consider incorporating the $s_2\ne 0$
term from Eq.~(\ref{sub}), which formally belongs to the two-loop order. 
We already mentioned above that perturbatively it evidently only affects the (re)-expanded
pressure (\ref{PP02lexp}) at order $\t m^4\lambda \sim \lambda^3$. Accordingly, it does not affect the perturbative order at which the moderate residual 
scale-dependence first reappears, Eq.~(\ref{2lmuremn}). However, 
the {\em nonperturbative} RG and OPT pressures are affected for larger coupling values, 
as intuitively expected since including $s_2\ne 0$ incorporates a 
(RG) part of the three-loop contributions. Indeed, as shown in Fig. \ref{P2loop_s2}, it
improves slightly the (nonperturbative) scale dependence of the RG pressure (long dashed), in addition
the OPT and RG pressures are closer to each other for large coupling values (but the scale dependence is still better for the RG solution than
for the OPT one).
More remarkably, with $s_2\ne 0$ the two-loop pressure obtained from the RG mass gap is seen to almost coincide with the one-loop pressure, up to relatively large
coupling $g\sim 1$, thus improving the apparent convergence further more.
Actually, this rather spectacular coincidence with the one-loop pressure for a large range of coupling values is partly accidental: 
after applying (\ref{subst1}), there are some partial cancellations of the two-loop contributions happening for relatively large coupling values (due 
to opposite signs $s_1, s_2$), between $s_1/3=-1/3 $ and the dominant two-loop term 
$s_2 \lambda = 24g^2 s_2$ (see (\ref{del2compact}), (\ref{sub}), (\ref{s0s1})), 
with a maximal cancellation for $g\simeq 1$, with $24g^2 s_2\simeq 1/3$. Since the $s_k$ values depend on the particular 
 RG coefficients, in a different theory $s_1$ and $s_2$ may have the same sign, or rather different magnitudes, possibly giving no such partial cancellations.
 Nevertheless the coincidence with the one-loop pressure is excellent even for relatively large intermediate coupling values {\it e.g.} $g\simeq 0.6-0.7$,  
 where $24g^2 s_2\simeq .12-.16$ does not make the cancellation
 much effective. Accordingly, there is also clearly 
 a more generic effective stabilization of the perturbative series resulting from the RGOPT construction, 
 with an improved convergence and scale dependence when incorporating higher RG order dependence, as intuitively expected.
\subsubsection{Comparison with other (SPT and HTLpt) variational resummation approaches}
This stability and improved scale-dependence 
is also illustrated in the next Fig. \ref{Prgopt_htlpt}, 
 where the RG pressure  is compared with a standard two-loop OPT/SPT~\cite{SPT3l,OPT3l}:  more precisely, discarding ${\cal E}_0$ in (\ref{sub}), 
taking $a=1/2$ in (\ref{subst1}), and using solely
Eq.~(\ref{OPT}). To compare with another mass prescription,  instead of the mass optimization we use the screening 
mass, Eq.~(\ref{mD1}) (consistently truncated at two-loop order),  
plugged in the expression of the free energy, similarly to 
the prescription mostly adopted for QCD HTLpt~\cite{HTLPT3loop}.~\footnote{Using the perturbative screening mass instead 
of the optimized mass gap is essentially the procedure in HLTpt applications to QCD  because the 
optimized mass has no real solutions~\cite{HTLPT3loop}}. 
One sees that using the optimized mass within the
SPT/OPT gives a better scale dependence, although SPT with optimized or screening mass both  
have a definitely stronger scale dependence than the RGOPT for moderately large coupling values.
\begin{figure}[!]
\epsfig{figure=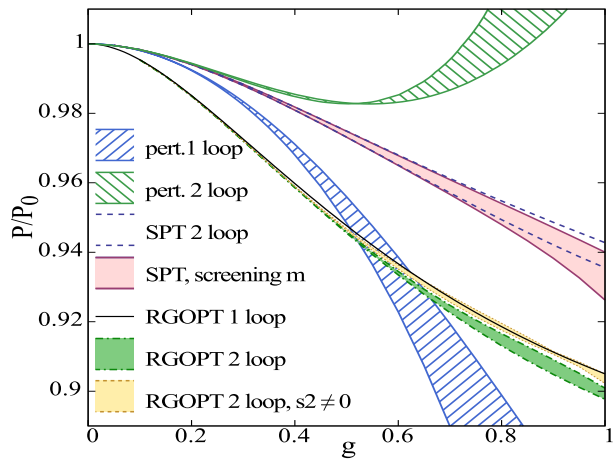,width=8cm,height=6cm}
\caption[long]{one- and two-loop RGOPT pressure versus one and two-loop standard perturbative pressure and two-loop  SPT 
pressure, with scale-dependence.}
\label{Prgopt_htlpt}
\end{figure}
To quantify what is illustrated in Fig. \ref{Prgopt_htlpt} more precisely, 
let us indicate the relative scale
variation of the various methods for the relatively large 
(rescaled) coupling value $g(2\pi T) =1$:  
the corresponding variation of $P/P_0$ between $\mu=\pi T$ and $4\pi T$
is $\simeq 8\%$, $0.8\%$, $1.5\%$, $0.35 \%$, and $0.3\%$ respectively 
for the 2-loop standard perturbation, SPT (optimized mass), SPT (screening mass), RGOPT, and RGOPT with $s_2\ne 0$.
Beyond $g\gsim 1$ the two-loop RGOPT scale-dependence increases more rapidly, but is still  only $\sim 1.6 \,10^{-2}$ 
for  $g\sim 1.5$ (while for such large coupling the relative 
variation of the standard perturbative two-loop pressure is as large
as $\sim 0.25$, and the scale variations of the HTLpt and standard OPT methods 
become very important too).  It would also be interesting to compare quantitatively our scale variation results 
with the residual scale-dependence appearing in the 2PI approach at three-loop order \cite{2PI3loop}.
However, a precise comparison appears difficult, since the renormalisation scheme and scale used 
in \cite{2PI3loop} are completely different and not easy to translate into the present scale variation 
in the $\ms$-scheme.\\
The RGOPT improvement on scale dependence at the two-loop order may not appear so spectacular 
as compared with SPT, merely a factor $\sim 2-4$ for $g\simeq 1$, essentially because the two-loop order 
SPT scale dependence here illustrated is still moderate. But at the three-loop level 
the SPT scale dependence becomes much larger~\cite{SPT3l} (and similarly HTLpt QCD\cite{HTLPT3loop}), of 
order $1$ for moderately large coupling. 
 As previously explained this can be traced both to $a\ne \gamma_0/b_0$ in (\ref{subst1}) together with the missing RG invariance 
from $ m^4 \ln\mu$ terms for arbitrary $m$ in (\ref{F02l}): since perturbatively $ m^4 \ln\mu \sim \lambda^2 \ln \mu $, the 
lack of scale invariance at formally one-loop order in the original massive free energy remains somewhat screened at 
one- and two-loop thermal perturbative expansion order, until it plainly  
resurfaces at three-loop $\lambda^2$ order. Remarkably the SPT pressure has even been calculated 
more recently to four-loop order~\cite{SPT4l}, and these   
results show very good convergence with respect to two- and thee-loop SPT for the central scale choice $\mu=2\pi T$. But
the remnant scale dependence is not illustrated in ~\cite{SPT4l}.
According to our general arguments  we do not see how the missing RG-consistent one-loop terms could be
compensated by going to higher orders if not present from the beginning. In contrast the RGOPT scale dependence remains very moderate, 
as illustrated in Fig. \ref{Prgopt_htlpt} when including a RG part of the three-loop contributions from $s_2\ne 0$. 
Moreover without explicitly calculating the full three-loop or higher order RGOPT pressure, we can be confident that the scale-dependence
 should further improve at higher orders: being built on 
perturbative RG invariance at order $k$ for {\em arbitrary} $m$, the RGOPT mass gap will exhibit remnant scale dependence as 
$\t m^2 \sim \lambda T^2 (1+\cdots+{\cal O}(\lambda^{k+1}\ln\mu))$, thus the dominant scale dependence in the free energy (namely the lowest perturbative
order at which scale dependence will resurface), 
coming from the leading term $-s_0\, m^4/\lambda$, should be of order ${\cal O}(\lambda^{k+2})$.  \\
Coming back to the two-loop order pressure, we have also checked the variation of our results against various perturbative truncations: 
for instance given that all our calculation relies
on the basic two-loop free energy (\ref{F02l}), it may be unnecessary refinement to use
the exact running coupling as in (\ref{runex2l}). Thus we looked for variation when
truncating the running Eq.~(\ref{run2lapprox}), at order $\lambda^2$, or $\lambda^3$ only. Also,
we studied the effect of truncating in the RG Eqs.~(\ref{RG2l}) and (\ref{RG2lfull}) terms 
of order $\lambda^3$, originating from the $b_1\lambda^3 \partial_\lambda$ 
term in (\ref{beta}), as it is formally one order higher than perturbatively 
strictly required (and the $b_1$-dependence enters
anyway at lower orders in the RG Eq. due to the $-s_0/\lambda$ subtraction). 
We obtain very good stability, since the maximal resulting variations, for the relatively
large coupling $g=1$, and scale choice $\mu=4\pi T$, is below $10^{-3}$ for $P_{\rm RG}/P_0$,
and somewhat worst but reasonably below $10^{-2}$ for $P_{\rm OPT}/P_0$.\\
\subsubsection{Full RGOPT two-loop solution}
Finally,  we can calculate $P^{(2)}_{\rm RGOPT}/P_0$ for the complete two-loop RGOPT 
solution, given by the non-trivial intersection between the RG and OPT equations (as illustrated in Fig. \ref{Mrgopt2Tall}),  
as function of the only free scale parameter, that we choose
as $t\equiv \mu/(2\pi T)$ (so the standard central scale choice corresponds to $t=1$). 
 Ideally for an all order calculation with exact scale invariance one would expect both  RG and OPT 
equations to give identical $\t m(\lambda,\mu/T)$ solutions. This is indeed the case
for simpler models, like typically~\cite{rgopt1} the large $N$-limit of the $O(N)$ GN-model.
But for a more involved theory at a finite order one expects
some differences between the RG and OPT solutions
since RG properties are only imposed perturbatively, those remnant differences reflecting 
the lack of exact RG/scale invariance. Just like the stationary mass solution is expected to approximate
the actually massless result, the intersection between the OPT and RG curves at a given order, defining
(variational) ``fixed-point'' mass and coupling, 
is expected to give a sensible approximation to the exactly scale-invariant nonperturbative result. \\
For the standard central scale choice $t=1$, the solution corresponding to the unique intersection of the physical RG and OPT branch
solutions, readily seen for $\t m$ and $\t g$ in 
Fig. \ref{Mrgopt2Tall}, gives:
\be
\t x\equiv \frac{\t m}{T} \simeq 0.912;\;\;\t g\equiv \sqrt{\frac{\t \lambda}{24}}\simeq 0.825;\;
 \frac{P^{(2)}_{\rm RGOPT}}{P_0}\simeq 0.907\;,
\label{fullrgopt}
\ee
obtained using the simpler high-$T$ expansion solutions. (NB for $t \gsim .5 $ approximately, 
$x(t)\equiv \t m(\mu)/T$ remains smaller than $1$, justifying a posteriori the high-$T$ approximation.
Typically, for $x=.9$ the high-$T$ approximation is already correct at the level
of $0.2\%$ and differences are completely negligible for smaller $x$.)\\ 
 The result in (\ref{fullrgopt}) needs further comments on its physical interpretation. 
Recall that the truly nonperturbative massless pressure expression $P(\lambda(\mu/T),\mu/T)/P_0 $, if that was available, would actually
be a function of the single coupling $\lambda(\mu_0=2\pi T_0)$ given at some input scale, thanks to exact scale invariance: for any renormalization 
scale choice $\mu$ the nonperturbative running $\lambda(\mu)$ would exactly compensate the explicit $\mu/T$ dependence. Incidentally, this is 
precisely the situation happening at the one-loop RGOPT order, where the exactly scale invariant $P/P_0$ in (\ref{P1P0G}) 
only depends on the single parameter $b_0\lambda(\mu_0)$. 
Now, at higher loop orders in the standard perturbative approaches (or similarly in the SPT/HTLpt approaches
after expressing the mass gap $\t m(\lambda)$ in terms of the coupling), due to imperfect scale invariance giving 
remnant perturbative terms $\ln^p(\mu/(2\pi T))$, 
one avoids large logarithms by fixing $\mu$ of order $\sim 2\pi T$,
which makes the scale effectively $T$-dependent and allows to study $P/P_0(\lambda(T/T_0))$ by varying the coupling as a function of the scale/temperature. 
In our case, using the second RG constraint Eq.~(\ref{RG}) enforces RG invariance at a limited (here two-loop) perturbative order, 
mimicking exact RG invariance. At $T=0$ two-loop order, as already mentioned in Sec. VII.A 
there is also a non-trivial intersecting OPT and RG solution at $\t g\sim 2.08, \ln (\t m/\mu)\simeq 0.08$ (see also Fig. \ref{LoptLrg}), 
that gives the free energy as ${\cal F}_0\propto \mu^4$, which still requires a reference physical scale
$\mu$ to be fully determined~\footnote{Similarly for QCD at zero temperature\cite{rgopt_Lam,rgopt_alphas}  the pion
decay constant has been used as a reference physical scale, and 
the analog of the combined RG and OPT solutions completely fixed 
$F_\pi/\Lambda_{QCD}$.}. At $T\ne 0$, 
combining the OPT and RG equation fixes $\lambda(\mu/T)$ and $m(\mu/T)$ for a given $\mu/T$ input, which also fixes $P/P_0$ 
as in (\ref{fullrgopt}). But this combined solution is somewhat academic for the $\phi^4$ model, 
specially in the symmetric phase  studied here, where there is no particular physical input temperature to compare with.
Moreover, since the scale invariance of the RGOPT pressure is still imperfect at two-loop order, the  
remnant scale dependence implies different $P/P_0$ values for different $\mu/T$ input choices. For instance, solving  
similarly the OPT and RG equations for $t=\mu/(2\pi T)=1/2$ (respectively $t=2$) gives $P/P_0 = 0.881$ (respectively $0.921$). 
This is consistent with the previously examined remnant scale dependence of the RG and OPT pressures at such relatively large coupling values. \\
As above mentioned one may expect that the full RGOPT solution 
for arbitrary scale input could give a sensible approximation of the genuine nonperturbative scale dependence of the coupling.
What Eq.~(\ref{fullrgopt}) indicates is that, for the physically reasonable
scale choice $\mu\sim 2\pi T$, the optimized coupling 
$\t g$ 	and mass $\t m/T$ are both of order $1$, lying in the rather nonperturbative range  where
the soft modes $\sim g T$ become comparable with the hard modes $\sim T$. Had we rather found
optimized values $\t x, \t g \ll 1$, we would not learn much beyond standard
perturbation theory. 
However, we cannot easily follow the combined solution over an arbitrarily large range of scale:  
the physical branch solution remains real for relatively large variations of $t>1$, but becomes complex
for $t \lsim 0.27$, which corresponds to the already mentioned complex RG solution for the large coupling 
$\lambda =-3b_0/(2b_1)$. 
Thus as already stressed we should not trust our results for $g$  above $g\gsim 1$, which corresponds
to $t \lsim .5$. 
For $t>1$ (respectively $t<1$) within a reasonable range, the combined RGOPT solution gives slightly smaller (respectively larger) 
optimized $\t g$ values as compared with the one in (\ref{fullrgopt}), unlike the standard perturbative RG behavior 
of the $\phi^4$ model at small coupling. Indeed for even larger $\mu \gg 2\pi T$, where one expects
to recover the four-dimensional $T=0$ $\phi^4$ model properties, the RGOPT real solution gives a 
slowly  (logarithmically) decreasing coupling and mass, which appears roughly consistent naively with the expected triviality property\cite{triviality}.   
Conversely, for $\mu\ll 2\pi T$  one expects to recover the
three-dimensional high-$T$ limit, such that after eventually reaching a maximum, the optimized coupling is expected to decrease again
for $\mu\ll 2\pi T$. But it is difficult to follow the full RGOPT solution becoming complex for $\mu\ll 2\pi T$. 
Nevertheless, even if they have no real intersections, the OPT and RG solutions can be reliably determined for $\mu\ll 2\pi T$,
as discussed in Sec. VII.B, and for fixed $m \ll 2\pi T$, $\lambda_{\rm OPT}(\mu/T)$ or $\lambda_{\rm RG}(\mu/T)$ 
decrease logarithmically for $\mu\ll 2\pi T$. \\
  Note that when incorporating the $s_2\ne 0$
term from Eq.~(\ref{sub}), the full RGOPT solutions similar to (\ref{fullrgopt}) are shifted to a different scale but with the same qualitative
behavior, so the net effect of $s_2\ne 0$ appears essentially as a renormalization scheme redefinition, without drastically changing
the results. Incidentally, for $s_2\ne 0$ the full RGOPT solution is already complex 
for $t=1$, while real solutions appear for slightly larger $t\gsim 1.5$ values, with 
corresponding optimal $P/P_0$ values very close to the one in (\ref{fullrgopt}). This simply reflects that 
the occurrence of complex RGOPT solutions in a given theory depend much on the 
renormalization scheme, so that real solutions could be recovered in principle from appropriate
perturbative scheme changes~\cite{rgopt_alphas}. But this is a much more involved task in the present finite temperature case. 
Moreover, since the non-trivial RG and OPT intersecting solution happens first at two-loop order, 
it is probably safer not to take as a very firm prediction the result in (\ref{fullrgopt}), nor 
to vary in a wide range around the preferred value $\mu\sim 2\pi T$. From the example of $T=0$ results in other 
models~\cite{rgopt_alphas}
the RGOPT variational fixed point solution is more likely to stabilize at the three-loop order, and probably with a more
perturbative value of the optimized coupling. 
\section{Conclusions and prospects}
Let us summarize our main results. We have shown that the 
standard treatment of the free energy (pressure) in massive thermal theories, with minimally 
subtracted counterterms, as primarily done in resummation approaches like OPT/SPT,HTLpt typically, 
lacks RG invariance. We have then recalled a general simple 
recipe to restore RG invariance,  leading unavoidably to
additional finite, temperature-independent vacuum energy contributions, systematically derivable in a perturbative fashion. 
We have next explained that the OPT/SPT,HTLpt resummation methods based on the 
modification of the perturbative expansion from the linear $\delta$-expansion in general do not preserve RG invariance,
which can however be restored  for a different interpolating prescription, Eq.~(\ref{subst1}), uniquely
dictated by universal first order RG coefficients, $a=\gamma_0/b_0$. Moreover, the resulting RGOPT 
resummation can use the RG equation as an alternative or additional combined constraint 
to determine a self-consistent nonperturbative mass (and coupling), in addition to the sole standard OPT prescription  
using only the mass optimization. We have then illustrated the RGOPT in details  by evaluating the free energy of a 
thermal scalar field theory  at one- and two-loop 
level. The results show a substantial improvement regarding this type of nonperturbative approach. Namely, we obtain exact RG/scale invariance
at the first non trivial one-loop RGOPT order, which also reproduces all the exact large-$N$ results~\cite{phi4N} of the $O(N)$ scalar model. 
At two-loop RGOPT order, the scale dependence and stability are 
drastically improved up to relatively large coupling values, with respect to most of the other available resummation approaches.
Not surprisingly the pressure obtained from the RG mass gap equation happens to have better convergence and scale dependence properties
for large coupling  than the pressure obtained from the OPT mass gap. We have also illustrated the full RGOPT result obtained from
combining the RG and OPT equations, therefore completely fixing the coupling and mass for a given input scale, with results that can be 
considered as a variational approximation to the truly scale invariant nonperturbative all order result.
Beyond two-loop order, since all relevant perturbative results are available at three-loop order~\cite{SPT3l},
the RGOPT procedure can be applied, but we leave this for future work as it becomes algebraically and numerically 
somewhat more involved. Besides, we are confident that it will further improve the scale dependence with respect to the two-loop results,
without explicit three-loop calculations:   
as explained the RGOPT construction will guarantee the RGOPT free energy to be perturbatively RG invariant up to neglected 
four-loop ${\cal O}(m^4\lambda^3)$ terms, for {\em arbitrary} mass $m$. This implies, after using the (OPT or RG) gap equation, 
that perturbatively a remnant scale dependence appears in the free energy at order $\sim \lambda^4$. 
The same line of reasoning also explains why the lack of RG invariance of OPT/SPT at formally one-loop order $m^4$ for arbitrary $m$, 
remains partly hidden at one- and two-loop but resurfaces maximally at perturbative three-loop order $\lambda^2$. We also see that the RGOPT 
anticipates the predictions by one perturbative order, the exact one-loop results being qualitatively 
similar to standard two-loop resummation results. Therefore, one may argue similarly that the
two-loop RGOPT results should be qualitatively comparable to (standard) SPT three-loop results~\footnote{Of course with the limitations that our two-loop
results, while resumming some RG dependence of higher order terms, does not reproduce the 
{\em full} three-loop contributions, in particular those given by the  three-loop 
thermal ``basketball'' graph~\cite{basketball,SPT3l}. When   
the subtraction term $s_2\lambda $ from (\ref{sub}), (\ref{s0s1}) is included, it 
is related from RG properties to the single logarithm coefficient of 
this three-loop basketball graph.}   
 (with a sensibly better scale invariance). Indeed, 
incorporating the $s_2$ subtraction term at two-loop order, which includes a (RG) part of the three-loop contributions, further improves
the convergence and scale dependence in the nonperturbative coupling range. 
For these reasons considering the full three-loop contributions is certainly a welcome refinement
but not necessary a crucial one in order to demonstrate the efficiency of the method which constitutes our main goal here. \\

It should be also clear  that 
the whole construction illustrated in particular in Secs. 3 to 5 is actually more
general, and that it is applicable to QCD. We have already mentioned some 
properties anticipated to be similar, or sometimes different, in the QCD case.
 One could expect that the thermal QCD application may be a priori much more difficult than the traditionally simpler scalar model.
But given that the difficult gauge dependence and related QCD issues have been solved by the HTL formalism~\cite{HTL,htlpt1}, the elaborate 
perturbative HTLpt calculations performed for thermal QCD up to three-loop order in \cite{HTLPT3loop} should be readily adaptable to our 
RGOPT method, which in a first stage relies entirely on perturbative calculations. 
In HTLpt only vacuum energy, mass, and coupling counterterms are necessary to renormalize the thermodynamic potential. The
quark mass anomalous dimension is just the standard one, while the gluon mass anomalous dimension is easily extracted from corresponding known counterterms, 
given e.g. in \cite{htlpt1,HTLPT3loop}. Thus our subtraction procedure to recover RG invariance will work just like in the scalar model, applying RG with
Eq.~(\ref{RGop}), (\ref{gamma}), and the modifed interpolation Eq.~(\ref{subst1}).
Moreover, the HTLpt formalism is inherently a high-$T$
expansion, therefore it will give OPT and RG equations in $m/T$ and $g$, simpler than the somewhat involved two-loop exact
$T$-dependent mass gap Eqs.~(\ref{OPT2l}), (\ref{RG2l}) (except that for QCD, HTLpt at two-loop order involves $m \ln m$ terms, 
due to the two-loop QCD free energy graph structure). Finally, for QCD the known first four coefficients of the $\beta$-function, $b_0\cdots b_3$, 
all have the same (negative) sign, such that no fake perturbative fixed points, such as the one present here in the scalar model, 
will obstruct the identification of correct RGOPT solutions. 
We are confident that a similarly improved scale-dependence and overall stability 
 will be obtained also from RGOPT adaptation of HTLpt, which could potentially put 
confrontation with thermal QCD lattice results on even firmer grounds. 
\acknowledgments
M.B.P. is partially supported by Conselho Nacional de Desenvolvimento Cient\'{\i}fico e Tecnol\'{o}gico (CNPq-Brazil).

\appendix
\section{Alternative derivation of subtraction terms from bare RG-invariance}
For completeness is this appendix we derive the subtraction
terms in Eq.~(\ref{s0s1}) alternatively by RG invariance considerations solely on the bare expression. 
The bare free energy at two-loop level
consists of Eq.(\ref{F02l})
plus the remnant 
divergent terms~\cite{vacanom_kastening,SPT3l} (after mass and coupling renormalization at this order)
with $D=4-2\epsilon$:
\be
(4\pi)^2 {\cal E}_0(\mbox{residual}) = -m^4 \left[ \frac{1}{4\epsilon} 
+\frac{1}{8\epsilon^2} \left (\frac{\lambda}{16\pi^2}\right ) \right]\;. 
\label{E0div}
\ee
As amply explained, minimally subtracting Eq.(\ref{E0div}) would not produce a finite 
RG-invariant expression. Instead, 
an explicitly RG-invariant counterterm can be written~\cite{gn2,qcd1,rgopt_alphas}  
in the general (perturbative) form in terms of the RG invariant bare mass and couplings as:
\be
(4\pi)^2 \Delta {\cal E}_0^{\rm RG} \equiv -\frac{m^4_0}{\lambda_0} H_0(\epsilon)\;\;,
\label{sub0}
\ee
with $H_0(\epsilon)\equiv \sum_{i\ge 0} h_i \epsilon^i$ 
an arbitrary series in $\epsilon$, most conveniently determined perturbatively order by order. 
Now demanding Eq.~(\ref{sub0}) to cancel the remnant 
divergent terms in (\ref{E0div}), upon
using in Eq.~(\ref{sub0}) the well-known expressions of the mass and coupling counterterms up to
two-loop order, reading in our convention:
\be
Z_\lambda \equiv \frac{\lambda_0}{\lambda} \simeq 1 +\frac{b_0}{2\epsilon}\lambda +
\left[\left (\frac{b_0}{2\epsilon}\right )^2 +\frac{b_1}{4\epsilon}\right] \lambda^2+\cdots
\label{Zlam2l}
\ee
\be
Z_m \equiv \frac {m_0}{m} \simeq 1 +\frac{\gamma_0}{2\epsilon}\lambda +
\left[\frac{\gamma_0(\gamma_0+b_0)}{8\epsilon^2} +\frac{\gamma_1}{4\epsilon}\right] \lambda^2+\cdots
\label{Zm2l}
\ee
expanded in $\lambda$ and $\epsilon$ series, 
after some algebra it uniquely determines $h_0=s_0$ and $h_1= s_1$, given in Eq.~(\ref{s0s1}).
But it leave additional 
finite subtraction terms identical to 
Eq.~(\ref{s0s1}). Note that it necessarily involves an $m^4 s_0/\lambda$: 
using in Eq.(\ref{sub0}) simply $m_0$ cannot cancel the one-loop divergence in Eq.(\ref{E0div})
since the latter  is ${\cal O}(\lambda^0)$. Of course for the present $\phi^4$ model this construction
is equivalent to the one performed in \cite{vacanom_kastening}.\\


\end{document}